\documentclass[10pt,aps,pra,twocolumn,floatfix,nofootinbib,superscriptaddress,longbibliography,nobalancelastpage]{revtex4-2}
\usepackage{placeins}
\usepackage{booktabs}
\usepackage{bm,graphicx,mathrsfs,amsmath,amssymb,mathtools,makecell,bbm, amsthm,dsfont,color,nicefrac,framed,enumitem,tikz,physics,wrapfig,amsfonts,tcolorbox,times,txfonts,lipsum}

\usepackage{comment}

\renewcommand{\ket}[1]{\lvert #1\rangle}
\renewcommand{\bra}[1]{\langle #1\rvert}

\renewcommand{\Tr}{\operatorname{Tr}}

\newtheorem{lemma}{Lemma}

\definecolor{alecolor}{RGB}{110,190,110}      
\definecolor{changescolor}{RGB}{0,120,0}

\definecolor{mycitecolor}{RGB}{204,170,0}     
\definecolor{mylinkcolor}{RGB}{110,190,110}   
\definecolor{myurlcolor}{RGB}{110,190,110}    

\usepackage[
    colorlinks=true,
    citecolor=mylinkcolor,
    linkcolor=mycitecolor,
    urlcolor=myurlcolor
]{hyperref}\usepackage{ragged2e}
\definecolor{equationcolor}{RGB}{154,135,198}
\usetikzlibrary{arrows.meta}
\usetikzlibrary{positioning}
\usetikzlibrary{calc}

\begin{document}
\title{Certification of non-stabilizerness in multi-qubit systems via worst-case random access codes}
\author{Santiago Zamora}
\affiliation{Department of Mathematics and Statistics, University of Ottawa, Ottawa, Ontario K1N 6N5, Canada}

\begin{abstract}
Recent works have proposed semi-device-independent frameworks to certify non-stabilizerness, but the complexity of the stabilizer polytope makes these methods infeasible beyond a few qubits.  In this work, we address this limitation by moving to partially characterized measurements in the context of worst-case random access codes. We first establish a simple condition, which we term the stabilizer pigeonhole criterion, that certifies non-stabilizerness in $m$-qubit pure states, and show how it can be applied to reduced states. As an application, we certify magic in $W$ states of arbitrary size using only eight preparations and three single-qubit measurement settings. Motivated by recent work on the stabilizer marginal problem, we then construct prepare-and-measure correlations generated by global $m$-qubit quantum states for which every $k$-qubit marginal lies inside the stabilizer polytope, yet the combined correlations cannot be reproduced by any global stabilizer mixture. Finally, we derive analytical upper and lower bounds on the worst-case bias attainable with mixed stabilizer preparations in an $m$-qubit random access code with  $n$ Pauli measurements. For the complete set of $4^m-1$ non-identity Pauli measurements, we show that the optimal quantum and mixed-stabilizer biases both scale as $2^{-3m/2}$, and that their ratio is asymptotically bounded by a constant ($\lesssim 1.85$), showing that in this setting, magic in the preparations alone yields at most a constant-factor advantage over stabilizer mixtures.
\end{abstract}

\maketitle
\section{Introduction}
Non-stabilizerness, often referred to as magic, is a fundamental resource in quantum computation \cite{Bravyi_2005, Veitch_2014}. Stabilizer circuits can be efficiently simulated on a classical computer~\cite{gottesman1997, gottesman1998}, and it is magic that elevates them to a regime of true quantum computational advantage~\cite{Bravyi_2005, Howard2014}. However, for quantum computers to eventually outperform classical hardware, it is not sufficient to inject magic into a few isolated qubits; we must understand how non-stabilizerness manifests itself in large-scale, many-body quantum systems. Consequently, quantifying and characterizing magic in many-body quantum states has become an important challenge \cite{Leone2022, Oliveiro_2022, Liu_2022}. 

Traditional characterization methods, such as full quantum state tomography, scale exponentially with the system size, rendering them infeasible to perform experimentally in the many-body regime. To bypass these experimental limitations, recent works have proposed semi-device-independent frameworks that trade highly demanding state reconstruction techniques for more accessible communication assumptions~\cite{Macedo_2025, Zamora2025}. Furthermore, as a byproduct of these investigations, it has become evident that magic is not only a computational resource but also a useful asset in communication protocols. For instance, non-stabilizerness provides distinct advantages in quantum state discrimination~\cite{stratton2026, Zhu_2024} and random access codes~\cite{Zamora2025, stratton2026}. Further research has established the relevance of magic in communication complexity~\cite{Chowdhury_2026, Girish_2025} and randomness extraction~\cite{Vairogs_2025}. However, despite the generality of these frameworks, their dependence on the geometry of the stabilizer polytope still renders them infeasible for more than a few qubits. 

Noting this limitation, this work shifts from the semi-device-independent regime of Ref.~\cite{Zamora2025} (where only the dimension of the system is assumed) to a partially characterized framework. We show that by assuming structure in the measurements, worst-case random access codes (RAC) offer a route towards scalable certification of many-body magic. By treating non-stabilizerness as a resource for dense information retrieval, we demonstrate that one can witness magic without  a full characterization of the state for multi-qubit systems. Our work is composed of the following four main contributions.

First, we compare the worst-case RAC communication capacities across classical, pure stabilizer, mixed stabilizer, and general quantum states, establishing a simple counting witness for pure states that we term the stabilizer pigeonhole criterion. Notably, we show that pure stabilizer states cannot encode $n$ bits with a worst-case success probability above $1/2$ as soon as $n > L_m = O(m^2)$, compared to the classical limit $n > m$ and to the limit $n > 4^m-1$ that holds for arbitrary quantum states.

Second, since applying the pure-state criterion globally requires an infeasible number of measurements for large $m$, we extend it to $k$-qubit reduced states. Relying on the fact that reduced states of pure stabilizer states are normalized stabilizer-code projectors---and thus form a finite set---we reduce the required number of measurements to $O(k^2)$, independently of the total system size $m$. As an illustration, we certify the magic of $m$-qubit $W$-states ($m \ge 3$) using only eight preparations and three single-qubit measurements.   

Third, because this reduced-state criterion relies on pure stabilizer preparations having marginals  restricted to the finite set of stabilizer-code projectors, it naturally raises a more challenging question, motivated by the stabilizer marginal problem studied in Ref.~\cite{Varela_2026}: can a worst-case RAC certify magic in a global state when every accessible marginal lies inside the stabilizer polytope? We answer this affirmatively by constructing explicit $m$-qubit preparations and $k$-local, pairwise anti-commuting Pauli measurements such that every $k$-qubit marginal belongs to the corresponding stabilizer polytope, yet the combined worst-case RAC statistics cannot be reproduced by any global stabilizer mixture. We also analyze the robustness of this construction under depolarizing noise, obtaining a visibility threshold of $\nu > 1/\sqrt{n}$.  

Fourth, since sets of pairwise anti-commuting Pauli observables are limited in size (at most $2m + 1$ exist on $m$ qubits~\cite{Bonet-Monroig_2020}), we drop the anti-commutativity restriction and derive general analytical upper and lower bounds on the worst-case bias for mixed stabilizer preparations with $n$ arbitrary Pauli measurements. To benchmark these results against general quantum strategies under the same measurements, we derive a complementary quantum upper bound from Haar-random Pauli expectations, which also independently disproves the attainability of the conjectured bound in Ref.~\cite{Liabotro_2017} for $m \ge 7$ under Pauli measurements. For the complete family of $4^m - 1$ non-identity Pauli observables, we find that both the optimal quantum and mixed-stabilizer biases scale as $2^{-3m/2}$ and differ by at most a constant factor ($\lesssim 1.85$). Hence, in this setting, magic in the preparations alone cannot provide more than a constant-factor advantage over mixtures of stabilizer states.

The remainder of the paper is organized as follows. In Sec.~\ref{sec:preliminaries}, we review stabilizer states and worst-case random access codes. Sections~\ref{sec:pigeonholecrit},~\ref{sec:reduction_pigeonholecrit},~\ref{sec:Marginal_Certif}, and~\ref{sec:general_bounds} present each of the four results above in turn. Finally, we discuss the implications of our results and outline open questions in Sec.~\ref{sec:discussion}.

\section{Preliminaries}\label{sec:preliminaries}
In this section, we introduce the concepts and quantities relevant for this work. We first introduce magic states, followed by a discussion on random access codes.

\subsection{Stabilizer theory}
A first relevant definition in the theory of stabilizer states is the Pauli group for $m$-qubit systems, denoted by $\mathcal{P}_m$. For a single qubit ($m=1$), it is given by~\cite{qubitguide}
\begin{equation}
    \mathcal{P}_1 = \{ \alpha P \mid P \in \mathcal{P}, \, \alpha \in \{\pm 1, \pm i\} \},
\end{equation}
where $\mathcal{P} = \{ \mathbb{I}, \sigma_x, \sigma_y, \sigma_z \}$ is the set of $1$-qubit Pauli observables. 

For $m$-qubit systems, the group consists of all $m$-fold tensor products of the Pauli matrices $P =(P_1 \otimes \dots \otimes P_m)\in\mathcal{P}^{\otimes m}$ equipped with the same global phase factors:
\begin{equation}
    \mathcal{P}_m = \{ \alpha P \mid P \in \mathcal{P}^{\otimes m}, \, \alpha \in \{\pm 1, \pm i\} \}.
\end{equation}
Note that the Pauli group $\mathcal{P}_m$ has $4^{m+1}$ elements, while the set of Pauli observables $\mathcal{P}^{\otimes m}$ has $4^m$. 

There are two equivalent ways to introduce stabilizer states: by introducing Clifford unitaries, i.e., unitaries $C$ that map the Pauli group to itself under conjugation, $C\mathcal{P}_mC^\dagger=\mathcal{P}_m$, with stabilizer states being those of the form $C\ket{0}^{\otimes m}$, or by introducing stabilizer groups~\cite{qubitguide, NielsenChuand2010}. 

A maximal stabilizer group is an abelian subgroup $\mathcal{S}\subseteq \mathcal{P}_{m}$ that does not contain $-\mathbb{I} $ and has $2^m$ elements. The group is generated by $m$ independent, commuting operators in $\mathcal{P}_{m}$. For every such stabilizer subgroup $\mathcal{S}$, there is a unique (up to a global phase) state $\ket{\mathcal{S}}$, such that $s\ket{\mathcal{S}} =\ket{\mathcal{S}}$ for every $s\in\mathcal{S}$. The state $\ket{\mathcal{S}}$ is called a stabilizer state. 

It is known that there are 
\begin{equation}
    N_m=2^m\prod_{j=1}^{m}(2^j+1)\label{eq:num_of_Stabstates}
\end{equation}
maximal $m$-qubit stabilizer groups, and therefore $N_m$ pure $m$-qubit stabilizer states. Let us denote the set of pure $m$-qubit stabilizer states by
\begin{equation}
    \mathrm{STAB}_m = \{\ket{\mathcal{S}}\mid \mathcal{S} \text{ is a stabilizer group}\}.
\end{equation}
One can consider mixtures of stabilizer states belonging to the convex hull of the set of pure stabilizer states,
\begin{equation}
    \mathrm{STABP}_m = \operatorname{conv}(\mathrm{STAB}_m),
\end{equation}
which forms a polytope. In our work, we will refer to any state in $\mathrm{STABP}_m$ as a stabilizer state. If necessary, we will specify whether it is a mixture or a pure state corresponding to the vertices of the polytope.  

Remarkably, a well-known result shows that if we restrict to states inside $\mathrm{STABP}_m$, any quantum computation using Pauli measurements and Clifford unitaries can be efficiently simulated by classical circuits~\cite{gottesman1997,gottesman1998}. Therefore, one is generally interested in certifying that the underlying system state lies outside of the stabilizer polytope; that is, the state is magic or non-stabilizer. This task, however, is difficult given the complexity of the stabilizer polytope. In this work, we approach the problem of certifying non-stabilizerness in $m$-qubit systems via worst-case random access codes. 

\subsection{Worst-case random access codes}
A random access code is a task performed in a scenario composed of two parties, Alice and Bob. In this scenario, Alice receives a random input string denoted by $x=x_1\dots x_n\in\{0,1\}^n$, while Bob receives an input $y\in\{1,\dots,n\}$. Alice encodes her inputs in an $m$-bit system and sends it to Bob. The goal is for Bob to guess the bit $x_y$ as accurately as possible by decoding the message using some algorithm $D_y$. Note that this is a specific task in the more general prepare-and-measure scenario  framework~\cite{junior2026}.

In this work, we consider quantum random access codes (QRAC), which means that Alice's messages are $m$-qubit systems $\{\rho_x\}_{x\in\{0,1\}^n}$ and Bob decodes the message by applying some general POVM $\{M_{0|y},M_{1|y}\}_{y=1}^n$. One can study different quantities in this protocol, the most common being the average success probability~\cite{junior2026}. However, for our purposes, it is more suitable to consider the worst-case probability~\cite{Tavakoli2020informationally}, defined as
\begin{equation}
    A(m,n) = \min_{x,y}p(b=x_y|x,y) =  \min_{x,y}\Tr(\rho_x M_{x_y|y}).\label{eq:worst-case-rac}
\end{equation}
This quantity represents the lowest probability of Bob correctly guessing the required bit $x_y$ across all possible inputs. Here $A(m, n)$ is the value of a given strategy; the optimal value over a class of strategies (classical, stabilizer, quantum) is denoted with a superscript, e.g., $A^C$ , $A^\mathrm{STAB}$, $A^Q$ below.

We can equivalently define the worst-case average, or bias, by
\begin{equation}
    \beta(m,n) = 2A(m,n)-1=2\min_{x,y}p(b=x_y|x,y)-1.
\end{equation}
For binary observables defined as
\begin{equation}
    B_{y} = M_{0|y}-M_{1|y}, \quad -\mathbb{I}\leq B_y\leq \mathbb{I} \quad \text{for}\quad y\in\{1,\dots,n\}, 
\end{equation}
and preparations $\{\rho_x\}_{x=1}^{2^n}$, the bias takes the form
\begin{equation}
    \beta(m,n) = \min_{x,y}\left[(-1)^{x_y}\langle B_y\rangle_{\rho_x}\right], \label{eq:bias}
\end{equation}
where $\langle B_y\rangle_{\rho_x} =\Tr(\rho_xB_y)$.

In what follows, we will use these quantities to study magic in $m$-qubit systems. We begin by studying the communication capacity in a RAC restricted to stabilizer states.

\section{Communication capacity of stabilizer states}\label{sec:pigeonholecrit}
In this section, we investigate how much information we can encode in a random access code using only stabilizer states. Establishing these bounds serves as the foundational baseline for our magic certification protocol. We divide our analysis into pure and mixed state encodings.
\subsection{Pure states}
First, consider the case of classical bit messages. There are $2^m$ possible messages in an $m$-bit system. These correspond to the basis states $C=\{\rho_x = \ket{x}\bra{x}\}_{x=1}^{2^m}$. Therefore, if there are $2^n$ inputs with $2^n>2^m$ (i.e., $n>m$), by the pigeonhole principle there necessarily exist two distinct labels $x\neq x'$ that correspond to the same state $\rho_x=\rho_{x'}$. Since $x\neq x'$, they differ in at least one entry such that $x_y\neq x'_y$. However, for these identical states, Bob performs the same measurement $M_y$. Therefore, the sum of his success probabilities is $\Tr(\rho_x M_{x_y|y}) + \Tr(\rho_{x'} M_{x'_y|y}) = 1 + \Tr((\rho_x-\rho_{x'}) M_{x_y|y})= 1$ using $M_{x'_y|y}=\mathbb{I}-M_{x_y|y}$ in the first equality and since $\rho_x=\rho_{x'}$. Because these probabilities sum to 1, they cannot both be strictly greater than $1/2$. Thus, for classical systems, the worst-case success probability is bounded:
\begin{equation}
    A^C(m,n) = \max_{\rho_x\in C, \, M_y\in \mathcal{M}}\min_{x,y}p(b=x_y|x,y) = \frac{1}{2},
\end{equation}
whenever $n>m$. $\mathcal M$ denotes the set of all measurements and we call $A^C(m,n)$ the classical value of $A$.  Therefore, observing a higher value $A(m,n) > 1/2$ in this regime certifies quantumness in the states under the assumption that the system has dimension $2^m$. 

An analogous derivation can be stated for the set of pure stabilizer states $\mathrm{STAB}_m$. Recall that the number of pure stabilizer states $N_m$ is given by Eq.~\eqref{eq:num_of_Stabstates}. Thus, we can define the pure stabilizer capacity (or pure stabilizer value) $A^\mathrm{STAB}(m,n)$ via random guessing:
\begin{equation}
    A^\mathrm{STAB}(m,n) = \max_{\rho_x\in \mathrm{STAB}_m, \, M_y\in \mathcal{M}}\min_{x,y}p(b=x_y|x,y) = \frac{1}{2},
\end{equation}
whenever $2^n>N_m$, or equivalently $n >\log_2 N_m$. Taking the floor $L_m=\lfloor\log_2 N_m\rfloor$ allows us to reduce the inequality to $n >L_m$, where
\begin{equation}
L_m=
\begin{cases}
2, & m=1,\\
5, & m=2,\\
\displaystyle\frac{m(m+3)}{2}+1, & m\geq3.\label{eq:Lm}
\end{cases}
\end{equation}
Notice then that, with respect to classical messages, using pure stabilizer states can increase the communication capacity of the $n$ inputs at most from $O(m)$ to $O(m^2)$. It does increase it at least to $2m$: for $m=1$, the stabilizer states $\{\pm\sigma_x,\pm\sigma_z\}$ and Bob's measurements $(\sigma_x\pm\sigma_z)/\sqrt2$ give $A=\frac12(1+1/\sqrt2)$ for $n=2>1$, and one can take products of such codes. Whether $L_m$ itself is attainable is not established here. Observing a value of $A(m,n) > 1/2$ while $n>L_m$ for pure preparations, then necessarily implies that at least one of the preparations has magic. We note that this is a sufficient witness: failure to observe a value above $1/2$ does not imply that the preparations are stabilizer states. We will refer to this criterion as the stabilizer pigeonhole criterion.  

As a simple example, consider the worst-case QRAC $A^{\mathrm{STAB}}(1,3)$. Here, Alice encodes $2^3 = 8$ bitstrings into $1$ qubit. Since $n=3 > L_1 = 2$, we have $A^S(1,3) = 1/2$. However, it is known that for general quantum strategies one can obtain~\cite{ambainis2009}
\begin{equation}
    A^Q(1,3) =\frac{1}{2}\left(1 + \frac{1}{\sqrt{3}}\right) \approx 0.788. \label{eq:max_31RAC}
\end{equation}

The optimal states are given by the Clifford orbit of the $T =  \frac{1}{2} \left( \mathbb{I} + \frac{\sigma_x + \sigma_y + \sigma_z}{\sqrt{3}}\right)$ state:
$\rho_x = \frac{1}{2} \left( \mathbb{I} + \frac{(-1)^{x_1}\sigma_x + (-1)^{x_2}\sigma_y + (-1)^{x_3}\sigma_z}{\sqrt{3}} \right)$, 
which are  the $T$-type (facet) magic states, i.e., the pure single-qubit states farthest from the stabilizer octahedron in the Bloch sphere.

\subsection{Mixed states}
Now assume that Alice can use local randomness, effectively mixing her stabilizer states. This allows her to have access to the full $m$-qubit stabilizer polytope $\mathrm{STABP}_m$. Notice now that the stabilizer polytope is a subset of the set of all quantum states. The affine dimension of the full quantum state space is exactly $4^m - 1$. By the dimension-counting argument of Refs.~\cite{Iwama_2007, Liabotro_2017}, it is impossible to extract a worst-case bias for $n > 4^m - 1$ bits using any quantum state. In particular, since $\mathrm{STABP}_m$ is a subset, it inherits this exact same impossibility bound.

On the other hand, the pure stabilizer states span the entire real vector space of Hermitian matrices. Therefore, their convex hull ($\mathrm{STABP}_m$) is full-dimensional within the quantum state space and it contains a neighborhood (a ball of radius $r > 0$) around the maximally mixed state $\mathbb{I}/2^m$. Since this neighborhood has the full dimension $4^m - 1$, one can trivially embed standard QRAC strategies~\cite{Liabotro_2017,Iwama_2007} (using appropriately scaled, orthogonal Pauli observables) entirely inside $\mathrm{STABP}_m$ for any $n \leq 4^m - 1$, leaving space for an advantage (explicitly, $\rho_x=\mathbb{I}/2^m+\epsilon\sum_{y}(-1)^{x_y}P_y$ with $\epsilon>0$ small enough and distinct non-identity Paulis $P_y$, measured by $P_y$).
In other words, the stabilizer value is
\begin{equation}
    A^{\mathrm{STABP}}(m,n) = \max_{\rho_{x}\in \mathrm{STABP}_m, \, M_y\in \mathcal{M}}\min_{x,y}p(b=x_y|x,y) > \frac{1}{2},
\end{equation}
if and only if $n \leq 4^m - 1$; the exact same relation holds for general mixed quantum states. For general \emph{pure} quantum states, $n\le 4^m-1$ is still necessary, but we do not show here that it is attainable (for fixed $x$, $\min_y p(b=x_y|x,y)$ is concave, not linear, in $\rho_x$, so optimal preparations need not be pure). It is interesting to compare this with the classical case, in which an advantage can be observed only if $n \leq 2^m - 1$~\cite{Liabotro_2017}.

Importantly, given the full dimension of the stabilizer polytope, one cannot derive an equivalent magic detection criterion based strictly on the number of bits $n$ as in the case of pure states. For pure states, just achieving $A>1/2$ at a high enough $n$ certifies magic. For mixed states, $n$ cannot certify magic because mixed stabilizers reach the exact same maximum input threshold ($4^m-1$) as arbitrary quantum states. To certify magic in the mixed case, one must look at the specific value of the success probability $A(m,n)$ (e.g., proving $A(m,n)$ exceeds the maximum threshold possible for the stabilizer polytope for fixed $m,n$). In Sec.~\ref{sec:general_bounds} we approach a more general case and bound the stabilizer value. Before that, in the following section we study the usefulness of the stabilizer pigeonhole criterion, while in Table~\ref{tab:encoding_thresholds} we summarize the encoding thresholds, including the stabilizer bounds found in this section. 

\begin{table}[htpb]
\centering
\renewcommand{\arraystretch}{1.5}
\begin{tabular}{@{}lcc@{}}
\toprule
\textbf{Framework} & \textbf{Pure States ($A > 1/2$)} & \textbf{Mixed States ($A > 1/2$)} \\ \midrule
\textbf{Classical}   & $n \le m$                             & $n \le 2^m - 1$                     \\
\textbf{Stabilizer} & $n \le \lfloor \log_2 N_m \rfloor =O(m^2)$$^\dagger$    & $n \le 4^m - 1$                     \\
\textbf{Quantum}   & $n \le 4^m - 1$$^\dagger$                       & $n \le 4^m - 1$                     \\ \bottomrule
\end{tabular}
\vspace{0.5em}
\caption{Upper bounds on the number of bits $n$ for which a worst-case success probability greater than $1/2$ can be achieved in an $m$-qubit system. For pure stabilizer states, the counting argument gives the necessary bound $n\leq\lfloor\log_2N_m\rfloor$, where $N_m=2^m\prod_{j=1}^m(2^j+1)$. For mixed-state encodings, the relevant threshold follows from the affine dimension of the state space. The daggered entries are necessary bounds whose attainability is not established in this work.}
\label{tab:encoding_thresholds}
\end{table}

\section{Stabilizer pigeonhole criterion for $k$-local measurements}\label{sec:reduction_pigeonholecrit}

Although the pure state criterion presented in the previous section is theoretically sound, it is difficult to implement in practice. Suppose one wants to certify the modest number of $m = 10$ qubits. To apply the criterion, Eq.~\eqref{eq:Lm} implies that we need $n > 66$. This requires more than $2^{66}$ preparations and $66$ measurements on all $m=10$ qubits, which is clearly infeasible. 

A first approach to reduce this number is to notice that any pure multi-qubit stabilizer state has reduced states that are also stabilizer states. Note that although they are mixed states, they form a specific finite set~\cite{Audenaert_2005}. Therefore, by restricting Bob to measuring a subset of $k$ qubits from the $m$-qubit system and applying the stabilizer pigeonhole criterion, one can aim to reduce the number of required measurements.  

\begin{figure}[htbp]
    \centering
    \includegraphics[width=0.9\linewidth]{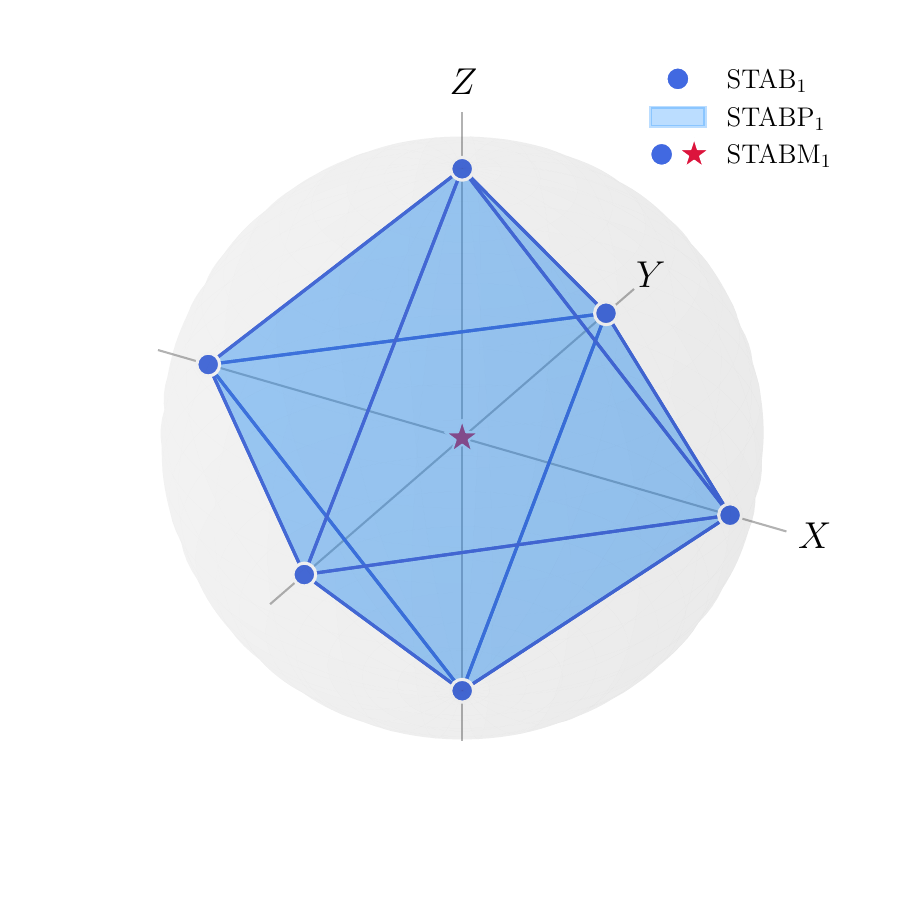}
    \caption{\textbf{The stabilizer polytope for a single qubit ($m=1$)}. The blue vertices ($\mathrm{STAB}_{1}$) represent the $6$ pure stabilizer states, whose convex hull is the octahedral polytope ($\mathrm{STABP}_{1}$). The red star at the origin is the maximally mixed state. Together with the vertices, it forms the set $\mathrm{STABM}_{1}$. The finite cardinality of $\mathrm{STAB}_{1}$ and $\mathrm{STABM}_{1}$ allows us to apply the stabilizer pigeonhole criterion}
    \label{fig:octahedron}
\end{figure}

We formalize the idea as follows. Consider $2^n$ pure preparations of $m$-qubits, denoted by $S_m =\{\ket{\psi_x}\}_{x=1}^{2^n}$. Assume first that all the preparations are stabilizer states, and define $S_k$ as the set of reduced states on the $k$-qubit subsystem of Bob's interest:
\begin{equation}
S_k = \left\{\rho_x^{(k)} = \operatorname{Tr}_{m-k}(|\psi_x\rangle\langle\psi_x|) \mid |\psi_x\rangle \in S_m \right\}.
\end{equation}
Although these reduced states are generally mixed, they are necessarily mixed stabilizer states, since partial tracing a pure stabilizer state yields a normalized projector onto a stabilizer code of the remaining subsystem. Consequently, $S_k$ is a subset of the set $\mathrm{STABM}_k$ of all $k$-qubit mixed stabilizer states (normalized stabilizer-code projectors; note that $\mathrm{STABM}_k\subset\mathrm{STABP}_k$), whose cardinality $N_k^{\mathrm{mix}}$ is~\cite{Gross2006,Singh2023}
\begin{equation}
    N_k^{\mathrm{mix}} = \sum_{r=0}^k 2^r \genfrac{[}{]}{0pt}{}{k}{r}_2 \prod_{i=0}^{r-1}(2^{k-i}+1).
\end{equation}
Here, the Gaussian binomial coefficient times the product counts the isotropic $r$-dimensional subspaces of $\mathbb{F}_2^{2k}$ (the stabilizer groups of rank $r$), while the factor $2^r$ accounts for the possible stabilizer sign assignments. (For $k=1,2$ this gives $N^{\mathrm{mix}}_1=7$ and $N^{\mathrm{mix}}_2=91$, see Fig.~\ref{fig:octahedron} for an example of this set in the $1$-qubit case.) 
Since this is still a finite, discrete number of states, we can apply the pigeonhole criterion directly to this set. The maximum worst-case success probability over these reduced states is bounded by random guessing:
\begin{equation}
    A^{\mathrm{STABM}_k}(m,n) = \max_{\rho_x^{(k)} \in \mathrm{STABM}_k, \, M_y \in \mathcal{M}} \min_{x,y} \Tr(\rho_x^{(k)}M_{x_y|y}^{(k)}) = \frac{1}{2},
\end{equation}
whenever $n > \log_2 N_k^{\mathrm{mix}}$. 

Therefore if we define
\begin{equation}
    A^{S_k}(m,n) =  \min_{x,y} \Tr(\rho_x^{(k)}M_{x_y|y}^{(k)}),
\end{equation}
the worst-case-probability of the reduced $k$-local scenario defined with preparations $S_k$ and fixed measurements $M^{(k)}_{b|y}$. Then if  $A^{S_k}(m,n)>1/2$ and $n > \log_2 N_k^{\mathrm{mix}}$, at least one of the reduced states in $S_k$ must lie outside the set $\mathrm{STABM}_k$ of normalized stabilizer-code projectors. Consequently, at least one of the underlying global pure preparations must be a non-stabilizer state.

One can further  consider the choice of the $k$-qubit subset on which Bob measures (the same subset for all $y$) and optimize over it, defining
\begin{equation}
    A(m,n,k) = \max_{S_k, M^{(k)}} \min_{x,y} \Tr(\rho_x^{(k)}M_{x_y|y}^{(k)}).
\end{equation}
Since the bound above holds for every fixed choice of subset, we can again conclude that if
\begin{equation}
     A(m,n,k) > \frac{1}{2} \quad \text{and} \quad n > \log_2 N_k^{\mathrm{mix}},
\end{equation}
then the underlying pure $m$-qubit preparations must contain magic. Finally, observe that this approach reduces the problem to $n =O(k^2)$, a value independent of the total number of qubits $m$.

To illustrate the reduction to local measurements, we consider an ensemble constructed from the $m$-qubit $W$-state,
\begin{equation}
    \ket{W_m} = \frac{1}{\sqrt{m}}\sum_{j=1}^m\ket{00\dots1_j\dots00}.
\end{equation}
We consider the multipartite cases $m\geq 3$. The key idea is to use Clifford operations for both the preprocessing and the encoding, ensuring that no magic is introduced by Alice's operations. 

As a common preprocessing operation, Alice applies a CNOT gate between the first and second qubits, with the first qubit as the control:
\begin{equation}
    C_{12}=\operatorname{CNOT}_{1\rightarrow 2}.
\end{equation}
The resulting state is $\ket{\phi_m}=C_{12}\ket{W_m}$. Focusing on the first qubit, its reduced state is
\begin{align}
    \rho_0^{(1)} &= \operatorname{Tr}_{2,\ldots,m} \left(C_{12}|W_m\rangle\langle W_m|C_{12}^\dagger\right) \nonumber \\
    &= \frac{1}{2}\left(\mathbb{I}+a_m\sigma_x+c_m\sigma_z\right). \label{eq:rho_0}
\end{align}
where we defined $a_m=2/m$ and $c_m=(m-2)/m$. Note that $a_m+c_m=1$, so $\rho_0^{(1)}$ lies on the boundary of the single-qubit stabilizer octahedron.

We now construct a three-bit random access code using only single-qubit Clifford encoding operations. First we define a unitary that rotates $\rho_0^{(1)}$ in $8$ different directions. For each three-bit string $x=(x_1,x_2,x_3)$, consider a single-qubit Clifford unitary $U_x$ satisfying
\begin{equation}
    U_x\sigma_x U_x^\dagger =
    \begin{cases}
    (-1)^{x_1}\sigma_x, & \text{if } x_1\neq x_2,\\
    (-1)^{x_1}\sigma_y, & \text{if } x_1 = x_2,
    \end{cases}
    \qquad
    U_x\sigma_z U_x^\dagger=(-1)^{x_3}\sigma_z.
\end{equation}
Such a Clifford exists for every $x$, since each assignment corresponds to a proper rotation of the octahedral group (the image of $\sigma_y=i\sigma_x\sigma_z$ is fixed by the other two).
Alice's final preparations $S_m = \{\ket{\psi_x}\}_{x\in\{0,1\}^3}$ are given by
\begin{equation}
    \ket{\psi_x} = U_x^{(1)}C_{12}\ket{W_m}, \qquad x\in\{0,1\}^3,
\end{equation}
where $U_x^{(1)}$ is the unitary that acts as $U_x$ on the first qubit and as the identity on all remaining qubits. Notice that since both $C_{12}$ and $U_x^{(1)}$ are Clifford operations, if the initial state $\ket{W_m}$ were a stabilizer state, all eight preparations would also be stabilizer states. The reduced states of these preparations, which form our target set $S_1 = \{\rho_x^{(1)}\}_{x\in\{0,1\}^3}$, are
\begin{equation}
    \rho_x^{(1)} =
    \begin{cases}
    \displaystyle
    \frac{1}{2}\left(\mathbb{I}+(-1)^{x_1}a_m\sigma_x+(-1)^{x_3}c_m\sigma_z\right), & x_1\neq x_2,\\[3mm]
    \displaystyle
    \frac{1}{2}\left(\mathbb{I}+(-1)^{x_1}a_m\sigma_y+(-1)^{x_3}c_m\sigma_z\right), & x_1 = x_2.
    \end{cases} \label{eq:rho_x}
\end{equation}
Bob performs three binary projective measurements on the first qubit, defined by
\begin{equation}
    M_{b|y}^{(1)} =\frac{1}{2} \left(\mathbb{I}+(-1)^b\vec{n}_y\cdot\vec{\sigma}\right),
\end{equation}
where
\begin{equation}
    \vec{n}_1=\frac{(1,1,0)}{\sqrt{2}}, \qquad \vec{n}_2=\frac{(-1,1,0)}{\sqrt{2}}, \qquad \vec{n}_3=(0,0,1).
\end{equation}
Using these observables and Eq.~\eqref{eq:rho_x}, we obtain
\begin{align}
    p(b=x_y|x,y) &= \operatorname{Tr}\left(\rho_x^{(1)}M^{(1)} _{x_y|y}\right) \nonumber \\
    &=
    \begin{cases}
    \displaystyle
    \frac{1}{2}\left(1+\frac{a_m}{\sqrt{2}}\right), & y=1,2,\\[3mm]
    \displaystyle
    \frac{1}{2}(1+c_m), & y=3.
    \end{cases}
\end{align}
Consequently, the worst-case success probability achieved by this strategy is
\begin{equation}
    A^{S_1}(m,3) = \frac{1}{2}\left[ 1+\min\left\{ \frac{\sqrt{2}}{m}, \frac{m-2}{m} \right\}\right]. \label{eq:advantage}
\end{equation}
In particular, this implies that
\begin{equation}
    A^{S_1}(m,3) > \frac{1}{2}, \qquad \text{for } m\geq 3.
\end{equation}

We can now apply the stabilizer pigeonhole criterion. Recall that the number of single-qubit mixed stabilizer states, in the sense of normalized stabilizer-code projectors, is $N_1^{\mathrm{mix}}=7$, corresponding to the six pure Pauli eigenstates and the maximally mixed state. Since our protocol uses eight preparations, we have
\begin{equation}
    n=3 > \log_2 N_1^{\mathrm{mix}} = \log_2 7 \approx 2.81.
\end{equation}
Therefore, if all eight global preparations were stabilizer states, their reduced states could not yield a worst-case RAC success probability exceeding $1/2$. However, Eq.~\eqref{eq:advantage} demonstrates a strict advantage over random guessing. It follows that at least one of the preparations must be a non-stabilizer state. Moreover, since all the generating operations are Clifford (no magic injection by Alice), the initial state $\ket{W_m}$ itself must be non-stabilizer for every $m\geq 3$.  

This example illustrates how the non-stabilizerness of an $m$-qubit resource state can be certified using only single-qubit measurements and eight preparations, independently of the total system size. For $m=10$, for instance, our protocol requires only eight preparations and three measurements on a single qubit---a drastic reduction from the exponentially large resources required by the original criterion. We stress what this certification assumes: the dimension of the system ($m$ qubits), that Bob's measurements act on the first qubit only, and that Alice's encoding consists of the specific Clifford operations above applied to a fixed initial state. Note also that the non-stabilizerness of $\ket{W_m}$ is not in question; the point of the example is to illustrate that the non-stabilizerness of $m$-qubit states can be certified, under the assumptions above, from single-qubit statistics, with a number of preparations and measurements independent of $m$. 

Also we note that the scalability here refers specifically to the number of preparations and measurement settings, not to a size-independent signal strength. Indeed, Eq.~\eqref{eq:advantage} gives a worst-case bias $\beta_W(m)=\min\{\sqrt{2}/m,(m-2)/m\}$, which equals $\sqrt{2}/m$ for $m\geq4$. Thus the protocol has constant setting complexity, while its statistical separation from random guessing becomes smaller as the system grows.

Importantly, although the individual reduced states in Eq.~\eqref{eq:rho_x} are convex mixtures of single-qubit stabilizer states, they are not among the seven normalized stabilizer-code projectors counted by $N_1^{\mathrm{mix}}$. This is what makes the criterion work: it uses that reduced states of \emph{pure} global stabilizer states can only belong to this finite set, which is a stronger constraint than membership in the stabilizer polytope. If the reduced states were only guaranteed to lie in $\mathrm{STABP}_k$, or if the global preparations were general mixed stabilizer states, the counting argument would break down. Can global magic still be certified when every accessible marginal belongs to the stabilizer polytope, such that no individual marginal independently reveals the non-stabilizerness? In the following section, we answer this affirmatively by considering incompatible stabilizer marginals and combining information from different subsystems within a single random access code experiment, an approach inspired by the stabilizer marginal problem of Ref.~\cite{Varela_2026}.

\section{Magic certification from incompatible stabilizer marginals.}\label{sec:Marginal_Certif}
\begin{figure}[htbp]
    \centering
    \includegraphics[width=0.95\linewidth]{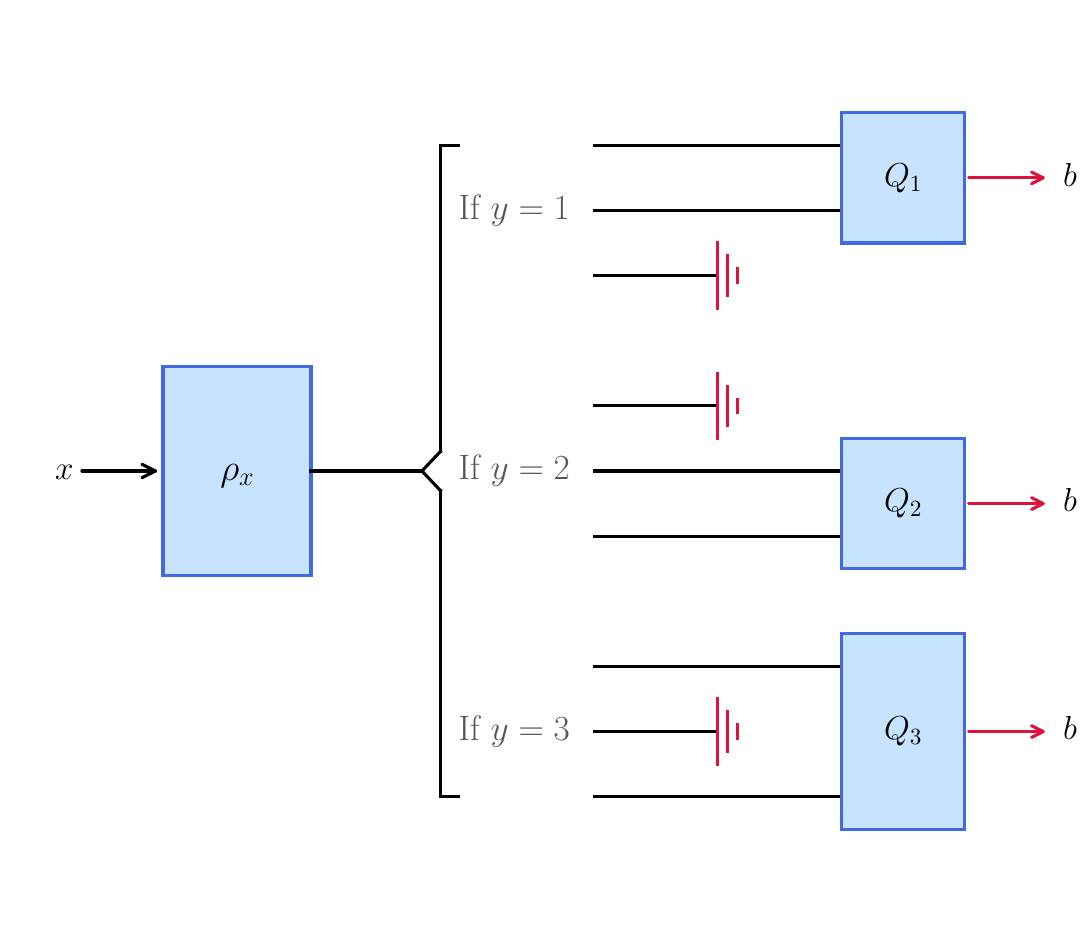}
    \caption{\textbf{A QRAC protocol for certifying global magic using incompatible marginals.} Alice prepares a $3$-qubit global state $\rho_x$ based on her classical input $x$. Depending on Bob's classical input $y \in \{1,2,3\}$, he performs a bipartite measurement ($Q_1$, $Q_2$, or $Q_3$) on a specific pair of qubits, tracing out the third unmeasured subsystem. While every two-qubit marginal belongs to the stabilizer polytope, combining the statistics of all three choices of $y$ can certify the global non-stabilizerness of the preparations.}
    \label{fig:qrac_cases}
\end{figure}

Consider the following QRAC. Assume Bob measures the observables (we use the notation $AB = A\otimes B$ for short) $Q_1=\sigma_x^1\sigma_x^2$, $Q_2=\sigma_y^2\sigma_y^3$  and 
$Q_3=\sigma_y^1\sigma_z^3$. That is, he acts only on two of the three qubits he receives at a time, namely on $\{1,2\}$, $\{2,3\}$, $\{1,3\}$.  
Note that these are pair-wise anti-commuting observables:
\begin{equation}
    \{Q_i,Q_j\} = 0, \quad\forall i\neq j\in\{1,2,3\}.
\end{equation}
Therefore, Bob's binary two-local measurement operators are given by
\begin{equation}
    M_{b|y} = \frac{1}{2}(\mathbb I + (-1)^bQ_y)\label{eq:BobsOp}
\end{equation}
Also, Alice prepares the eight 3-qubit messages in the states
\begin{equation}
    \rho_x = \frac{1}{8} (\mathbb I + \frac{1}{\sqrt{3}}\sum_{i=1}^3(-1)^{x_i}Q_i),\label{eq:3qubit_preps}
\end{equation}
for $x=(x_1,x_2,x_3)\in\{0,1\}^3$. See Fig.~\ref{fig:qrac_cases} for an illustration of the protocol.

With these definitions, one can show that $p(b=x_y|x,y) = (1+1/\sqrt{3})/2$ for all $x$ and $y$. By denoting $A_k(m,n)$ the random access code of $m$-qubit systems with $n$ $k$-local measurements (not to be confused with $A(m,n,k)$ of Sec.~\ref{sec:reduction_pigeonholecrit}), then we have 
\begin{equation}
    A_2(3,3) =  (1+1/\sqrt{3})/2\approx0.7886.
\end{equation}

On the other side, consider that Alice prepares stabilizer states $\{\tau_x\}_x$. In this case we have 
\begin{equation}
    \sum_{i=1}^3|\langle Q_i\rangle_{\tau_x}|\leq 1.\label{eq:bound}
\end{equation}
To see this, start with a pure state $\ket{\phi}$ and  suppose that $\sum_{i=1}^3|\langle Q_i\rangle_\phi|> 1$. Then since for a stabilizer state $\ket{\phi}$, $\langle Q_i\rangle_\phi\in\{0,1,-1\}$, this implies there exist at least two operators $Q_i$,$Q_j$ for $i\neq j$ such that $Q_i\ket{\phi}=\pm\ket{\phi}$ and  $Q_j\ket{\phi}=\pm\ket{\phi}$, say with eigenvalues $s_i,s_j\in\{\pm1\}$. Then $Q_iQ_j\ket{\phi}=s_is_j\ket{\phi}=Q_jQ_i\ket{\phi}$, so $\{Q_i,Q_j\}\ket{\phi}=2s_is_j\ket{\phi}\neq0$, which contradicts $\{Q_i,Q_j\}=0$.  It extends to any stabilizer mixture, since they are convex combinations of pure stabilizer states and the left-hand side of Eq.~\eqref{eq:bound} is convex.   

We note that the constraint in Eq.~\eqref{eq:bound} is precisely the geometric restriction associated with the projected stabilizer polytope for a set of three pairwise anticommuting Pauli observables, as characterized in Ref.~\cite{Varela_2026}. In this case, the corresponding projected polytope is the octahedron defined by $\sum_{i=1}^{3}|\langle Q_i\rangle|\leq 1$. This provides a geometric interpretation of the stabilizer bound above. Our goal here, however, is to exploit this constraint within a worst-case random access code, thereby obtaining an operational certification of global non-stabilizerness even when no individual marginal reveals it.

By using Eq.~\eqref{eq:BobsOp}, one can write 
\begin{align}
    p(b=x_y|x,y)  &=\frac{1}{2} +\frac{1}{2}(-1)^{x_y}\Tr(\tau_xQ_y)\\
    &\leq\frac{1}{2} +\frac{1}{2}|\langle Q_y\rangle_{\tau_x}|.
\end{align}
By  noting that $\min_y|\langle Q_y\rangle_{\tau_x}|\leq\frac{1}{3}\sum_{i=1}^3|\langle Q_i\rangle_{\tau_x}|$ and applying the bound Eq.~\eqref{eq:bound}, one obtains $\min_yp(b=x_y|x,y) \leq 2/3$, a bound that is saturated by the mixtures $\tau_x=\frac13\sum_{i=1}^3\tau_{i,x_i}$, where $\tau_{i,b}$ is a pure stabilizer state with $\langle Q_i\rangle=(-1)^b$ (it has $\langle Q_j\rangle=0$ for $j\neq i$ by anti-commutation). This immediately gives
\begin{equation}
    A^{\mathrm{STABP}}_2(3,3) = \frac{2}{3}.
\end{equation}
This certifies the non-stabilizerness of preparations~\eqref{eq:3qubit_preps} under the assumption of the specific measurements~\eqref{eq:BobsOp} and the dimension of the system. However, it demonstrates the following property. Observe that the reduced states 
\begin{align}
\rho_x^{12} &=\frac14\left(\mathbb I_{12}+\frac{(-1)^{x_1}}{\sqrt3}\sigma_x^1\sigma_x^2\right),\nonumber\\
\rho_x^{23} &=\frac14\left(\mathbb I_{23}+\frac{(-1)^{x_2}}{\sqrt3}\sigma_y^2\sigma_y^3\right),\\
\rho_x^{13} &=\frac14\left(\mathbb I_{13}+\frac{(-1)^{x_3}}{\sqrt3}\sigma_y
^1\sigma_z^3\right),\nonumber
\end{align}
all belong to the stabilizer polytope $\mathrm{STABP}_2$, and all single qubit marginals are maximally mixed. The construction therefore exhibits a separation between local and global stabilizer compatibility at the level of an operational random access code. Although every two-qubit marginal belongs to the corresponding stabilizer polytope, the global quantum preparations achieve a worst-case success probability of $A_2(3,3)=(1+1/\sqrt{3})/2\approx 0.7887$, whereas any global stabilizer-mixture strategy is bounded by $2/3$, a bound that is tight. Consequently, the combined measurement statistics certify the non-stabilizerness of the global preparations, even though no individual two-qubit marginal reveals it.

Even more, we can generalize the construction to $2^n$ preparations of $m$ qubits. Suppose we can find $n$-pairwise anticommuting  Pauli observables $\{Q_1,\dots,Q_n\}$ on an $m$-qubit system with distinct supports $S_j$, each of size $k$ ($k$-local Pauli operators). Define the preparations 
\begin{equation}
    \rho_x = \frac{1}{2^m} (\mathbb I + \frac{1}{\sqrt{n}}\sum_{i=1}^n(-1)^{x_i}Q_i), \label{eq:marg_preps_m}
\end{equation}
and Bob's measurements as in Eq.~\eqref{eq:BobsOp} for $y\in\{1,\dots,n\}$.
Note that every $k$-qubit marginal of $\rho_x$ is of the form $2^{-k}(\mathbb I+t\,Q_i)$ with $|t|\leq1/\sqrt{n}$, which belongs to $\mathrm{STABP}_k$. To see why, consider a marginal on a specific $k$-qubit subset $T$. A Pauli operator $Q_i$ with support $S_i$ survives the partial trace if and only if $S_i \subseteq T$. Since both $S_i$ and $T$ have size $k$, this requires $S_i = T$; because the supports are distinct, at most one operator survives. The resulting marginal is therefore of the form $2^{-k}(\mathbb{I} + t Q_i)$ with $t = \pm 1/\sqrt{n}$. Furthermore, since $Q_i$ is a Pauli operator, its eigenstates are pure stabilizer states, meaning this marginal is a valid convex mixture of pure stabilizer states and thus belongs to $\mathrm{STABP}_k$.

By following the same reasoning for the specific case of $n=3$, $m=3$, and $k=2$ as described above we can obtain 
\begin{equation}
    A_k(m,n) =\frac{1}{2}(1+\frac{1}{\sqrt{n}}).
\end{equation}
On the other hand, the restriction to stabilizer states again implies $\sum_{i=1}^n|\langle Q_i\rangle_{\tau_x}|\leq 1$, which by an analogous derivation as above, one reaches 
\begin{equation}
    A^{\mathrm{STABP}}_k(m,n) = \max_{\tau_x\in \mathrm{STABP}_m}\min_{x,y} p(b|x,y) = \frac{1}{2}(1+\frac{1}{n}),\label{eq: Stab_bound_marg}
\end{equation}
giving a gap $\Delta$ of
\begin{equation}
    \Delta =\frac{\sqrt n-1}{ 2n}>0.
\end{equation}
In summary, for any $m$-qubit system and $n\geq 2$ pairwise anticommuting $k$-local Pauli observables with distinct supports, this construction yields valid quantum preparations $\rho_x$ for which every $k$-qubit marginal lies inside the $k$-qubit stabilizer polytope. Despite this local stabilizer compatibility, the quantum strategy achieves a worst-case success probability of
$A_k(m,n)=\frac{1}{2}\left(1+\frac{1}{\sqrt{n}}\right)$. Conversely, any strategy using global stabilizer-mixture preparations and the same measurements is upper-bounded by $A_k^{\mathrm{STABP}}(m,n)\leq\frac{1}{2}\left(1+\frac{1}{n}\right)$. As a consequence, we obtain an operational gap $\Delta=\frac{\sqrt{n}-1}{2n}>0$, which certifies global non-stabilizerness even when every accessible $k$-qubit marginal lies inside the stabilizer polytope.

Furthermore, we can explore the robustness of the protocol by considering depolarizing noise. In this case, the state preparations are
\begin{equation}
\rho_x(\nu) = \nu\rho_x + (1-\nu)\frac{\mathbb{I}}{2^m},
\end{equation}
where $\rho_x$ is given by Eq.~\eqref{eq:marg_preps_m} and $\nu\in[0,1]$. Note that the reduced states are inside of the stabilizer polytope since they are just noisy versions of the marginals of the ideal states,~Eq.~\eqref{eq:marg_preps_m}. It can be shown that the worst-case success probability under the $n$ anti-commuting Paulis transforms into
\begin{equation}
A_k^\nu(m,n) = \frac{1}{2}\left(1 + \frac{\nu}{\sqrt{n}}\right).
\end{equation}
To successfully certify non-stabilizerness, this noisy value must exceed the stabilizer bound obtained in Eq.~\eqref{eq: Stab_bound_marg},
\begin{equation}
\frac{1}{2}\left(1 + \frac{\nu}{\sqrt{n}}\right) > \frac{1}{2}\left(1 + \frac{1}{n}\right).
\end{equation}
which implies a visibility threshold of
\begin{equation}
\nu > \frac{1}{\sqrt{n}}. 
\end{equation}
In the particular case of $n=3$, we have $\nu > 1/\sqrt 3 \approx 0.5774$, showing that the three-qubit construction tolerates a substantial amount of white noise under ideal measurements.

The global-marginal certification is conditional on the known system dimension, the specified supports $S_i$, and Bob's use of the calibrated binary Pauli measurements $M_{b|i}=(\mathbb I+(-1)^bQ_i)/2$. It is therefore an operational witness under these assumptions. The limitation of this set up, however,  is that the existence of sets of anti-commuting operators is not  guaranteed since at most $2m+1$ of them exist on $m$ qubits (necessarily $n \leq \min\{2m+1,\binom{m}{k}\}$, the second bound coming from the requirement of distinct supports), and finding scalable constructions is a problem in the literature~\cite{Sarkar2021}. In the next section we remove this condition and derive general bounds for the stabilizer value of any worst-case random access code with Pauli measurements.

\section{General bounds for worst-case random access codes}\label{sec:general_bounds}
As mentioned in the last section, it is not always possible to find pairwise anti-commuting Paulis, restricting the application of the method presented there. In this section we approach the problem from another point of view. We derive upper and lower bounds on the RAC~\eqref{eq:worst-case-rac} restricted to stabilizer states and Pauli measurements without the anti-commutativity assumption. While extending this framework to general POVMs is a valid mathematical pursuit (and in fact closer to a device-independent certification), the continuous parameterization of arbitrary measurements obscures the discrete geometric properties of the stabilizer polytope. Restricting to Pauli observables allows us to derive asymptotic analytical bounds, whereas arbitrary measurements would possibly require numerical methods that cannot easily yield scaling laws for large $m$.

In what follows let $d=2^m$ and $D=d^2-1=4^m-1$. We will work with the bias defined in Eq.~\eqref{eq:bias}.  We assume we have $m$-qubit stabilizer preparations $\{\rho_x\}_{x\in\{0,1\}^n}\subset\mathrm{STABP}_m$ and we restrict Bob's measurements to $n$ distinct (up to sign) non-identity Pauli observables $B_y\in \mathcal{P}^{\otimes m}\setminus\{\mathbb I\}$ (so $n\le D$; the sign can be absorbed in a relabeling of $x_y$, so we write $B_y=P_y$). Note that, for fixed $x$, $\min_y[(-1)^{x_y}\langle P_y\rangle_{\rho_x}]$ is concave but not linear in $\rho_x$, so the maximum is in general \emph{not} attained on pure stabilizer states: in Sec.~\ref{sec:Marginal_Certif}, pure stabilizer states give $\beta=0$ for $Q_1,Q_2,Q_3$ while a mixture gives $\beta=1/3$. We therefore keep mixed preparations and write
\begin{align}
    \beta^{\mathrm{STABP}}_{\mathrm{Pauli}}(m,n) =\max_{\rho_x\in\mathrm{STABP}_m}\min_{x,y}\left[(-1)^{x_y}\Tr(\rho_xP_y)\right] \label{eq:bias_Stab}
\end{align}
The bounds below only use that, for fixed $x$, the sum $\sum_y(-1)^{x_y}\Tr(\rho_xP_y)$ is linear in $\rho_x$, and hence is bounded by its maximum over pure stabilizer states. Our goal is to bound this value above and below  as tightly as possible, under the assumption in which Bob measures Pauli observables. 
\subsection{Upper bound}
First, we focus on the upper bound. It is useful to note that since $\beta(m,n) = \min_{x,y}[(-1)^{x_y}\Tr(\rho_xP_y)]$, for any $x,y$ we have that
\begin{equation}
    \beta(m,n) \leq (-1)^{x_y}\Tr(\rho_xP_y)\label{eq:useful_bound}
\end{equation}
Now we state the following well known lemma (the proof is in Appendix \ref{app:pauli_norm}).
\begin{lemma}[Pauli expectations of stabilizer states]\label{lemma:pauli_norm}
Any pure $m$-qubit stabilizer state $\ket{\phi}$  satisfies
\begin{equation}
    \sum_{P \in \mathcal{P}^{\otimes m} \setminus \{\mathbb I\}} |\bra{\phi}P\ket{\phi}| = d-1.
\end{equation}
\end{lemma}

The upper bound is composed of two parts. For the first one, we simply note that by taking the absolute value and summing over all $y\in\{1,\dots,n\}$, Eq.~\eqref{eq:useful_bound} gives 
\begin{equation}
    n\beta(m,n) \leq \sum_{y=1}^n|\Tr(\rho_xP_y)|.
\end{equation}
We can bound the RHS of the above inequality with Lemma~\ref{lemma:pauli_norm} (for mixed $\rho_x$, decompose it into pure stabilizer states and use the triangle inequality) obtaining 
\begin{equation}
    \beta_{\mathrm{Pauli}}^{\mathrm{STABP}}(m,n) \leq \frac{d-1}{n}.
\end{equation}
As one can see this bound is loose for large $m$ since we are bounding the average value over the $n$ observables of Bob with a value obtained over the full set of Pauli operators. Therefore for large $m$ we obtain a tighter bound (second part of the bound) as follows. 

Let us take the sum over $y$ in Eq.~\eqref{eq:useful_bound}, obtaining
\begin{equation}
    n\beta(m,n) \leq \sum_{y=1}^n(-1)^{x_y}\Tr(\rho_xP_y)\;\leq\; \max_{\phi\in\mathrm{STAB}_m}\sum_{y=1}^n(-1)^{x_y}\bra{\phi}P_y\ket{\phi}.\label{eq:beta_2ndbound}
\end{equation}
Now assume that the strings $x$ are obtained uniformly at random. Therefore we can  define the random variable 
\begin{equation}
    X_\phi(x) =\sum_{y=1}^n(-1)^{x_y}\bra{\phi}P_y\ket{\phi}=\sum_{y=1}^n\alpha_y.
\end{equation}
Since Eq.~\eqref{eq:beta_2ndbound} holds for every $x$, it also holds on average over $x$: we have $n\beta\le\min_x\max_\phi X_\phi(x)\le\mathbb{E}_x\max_\phi X_\phi(x)$. By applying Hoeffding's Lemma~\cite[p.~21]{massart2007} to each term in $X_\phi(x)$ and using Jensen's inequality, we can bound the average value $ \mathbb E_x (\max_\phi X_\phi(x))\leq\sqrt{2(d-1)\ln N_m}$, resulting in 
\begin{equation}
    n\beta_{\mathrm{Pauli}}^{\mathrm{STABP}}(m,n)\leq\sqrt{2(d-1)\ln N_m}.\label{eq:upperbound2}
\end{equation}
See Appendix~\ref{app:upper_bound} for a detailed derivation. We can combine both bounds together and write,
\begin{equation}
    \beta_{\mathrm{Pauli}}^{\mathrm{STABP}}(m,n)\leq\frac{1}{n}\min\left(2^m-1,\sqrt{2(2^m-1)\ln N_m}\right).
\end{equation}
In subsection~\ref{sec:assymptotic_behaviour}, we discuss the consequences of this bound. First, in the next subsection we study its tightness by constructing a lower bound. 
\subsection{Lower bound}
To lower bound Eq.~\eqref{eq:bias_Stab}, we construct an explicit stabilizer strategy. The idea behind this construction is to reproduce the optimal quantum strategy inside the stabilizer polytope. Define the preparations
\begin{equation}
    \rho_x = \frac{\mathbb I}{d} +\epsilon\sum_{y=1}^n(-1)^{x_y}P_y.
\end{equation}
Our goal is to find the highest $\epsilon$ while staying inside the stabilizer polytope.
Let us consider the Hilbert-Schmidt distance of each preparation to the center of the polytope $\mathbb I/d$
\begin{equation}
    \big\|\rho_x-\frac{\mathbb I}{d}\big\|_{HS}^2 = \Tr((\rho_x-\frac{\mathbb I}{d})^2) = \epsilon^2 dn.
\end{equation}
Since we want them to be inside the polytope but as closest as possible to the optimal quantum strategy, we ask for 
\begin{equation}
    \big\|\rho_x-\frac{\mathbb I}{d}\big\|_{HS}^2 = r_{in}^2,
\end{equation} where $r_{in}$ is the radius of the biggest Hilbert-Schmidt ball centered at $\mathbb I/d$ that is contained in the polytope. This implies that $\epsilon=\frac{r_{in}}{\sqrt{nd}}$. Substituting these states in Eq.~\eqref{eq:bias}, with $B_y=P_y\neq\mathbb I$ we get
\begin{equation}
    \beta(m,n) = r_{in}\sqrt{\frac{d}{n}},
\end{equation}
which implies that 
\begin{equation}
    \beta_{\mathrm{Pauli}}^{\mathrm {STABP}}(m,n)\geq r_{in}\sqrt{\frac{d}{n}}.
\end{equation}
To finish,  we use the recent result presented in Ref.~\cite{liu2026} (Theorem 1). It shows that the inradius is bounded by 
\begin{equation}
    r_{in}\geq\sqrt{\frac{a_*}{d(d-a_*)}}, \quad a_*\approx0.4583.
\end{equation}
This gives the lower bound of
\begin{equation}
    \beta^{\mathrm {STABP}}_{\mathrm{Pauli}}(m,n)\geq\sqrt{\frac{a_*}{n(d-a_*)}}.
\end{equation}
Putting the upper bound and lower bound together we obtain that
\begin{equation}
     \sqrt{\frac{a_*}{n(d-a_*)}}\leq\beta^{\mathrm {STABP}}_{\mathrm{Pauli}}(m,n)\leq\frac{1}{n}\min\left(d-1,\sqrt{2(d-1)\ln N_m}\right).
\end{equation}

\subsection{Asymptotic analysis and quantum benchmark}
\label{sec:assymptotic_behaviour}

Before discussing the asymptotic behaviour, it is useful to illustrate that the stabilizer upper bound derived above can indeed be violated by general quantum strategies, even when Bob is restricted to Pauli measurements. Consider the one-qubit, three-bit QRAC, with $P_1=\sigma_x$, $P_2=\sigma_y$, and $P_3=\sigma_z$. For $m=1$ and $n=3$, we have $d=2$ and $N_1=6$. Our stabilizer upper bound therefore gives
\begin{equation}
    \beta^{\mathrm{STABP}}_{\mathrm{Pauli}}(1,3)  \leq \frac{1}{3} \min\left\{1,\sqrt{2\ln 6}\right\}=\frac{1}{3}.
\end{equation}
This bound is tight, as can be seen from the stabilizer mixtures
\begin{equation}
    \rho_x=\frac{1}{3}\sum_{j=1}^{3}\frac{1}{2}\left(\mathbb I+(-1)^{x_j}\sigma_j\right),
\end{equation}
which achieve a bias of $1/3$. On the other hand, as we mentioned in Sec.~\ref{sec:pigeonholecrit}, $T$-type magic states achieve the value
\begin{equation}
    \beta^Q(1,3) = \beta_{\mathrm{Pauli}}^Q(1,3)=\frac{1}{\sqrt{3}} >\frac{1}{3}=\beta^{\mathrm{STABP}}_{\mathrm{Pauli}}(1,3).
\end{equation}
In fact, the corresponding (optimal) decoding strategy  corresponds to measuring the three Pauli observables~\cite{ambainis2009}.  Therefore, this example illustrates that the stabilizer upper bound provides a nontrivial certification threshold even when Bob is restricted to Pauli measurements.

We now consider the asymptotic behaviour for large $m$, focusing on the case in which Bob measures all non-identity Pauli observables. The general stabilizer upper bound derived above gives, for large  $m\geq5$,
\begin{align}
    \beta^{\mathrm{STABP}}_{\mathrm{Pauli}}(m,D)&\leq\frac{\sqrt{2(d-1)\ln N_m}}{D} \nonumber\\
    &\approx\frac{m\sqrt{\ln 2}}{d^{3/2}},\label{eq:scaling_stab_Bound}
\end{align}
where we have used
\begin{equation}
    \ln N_m
    =\frac{m(m+3)}{2}\ln 2+O(1).
\end{equation}
On the other hand, the inradius construction gives
\begin{equation}
    \beta^{\mathrm{STABP}}_{\mathrm{Pauli}}(m,D)\geq \sqrt{\frac{a_\ast}{D(d-a_\ast)}}\approx
    \frac{\sqrt{a_\ast}}{d^{3/2}}.\label{eq:scaling_stab_lower}
\end{equation}
Thus, these two stabilizer bounds have the same exponential scaling, differing by a factor of order $m$. We next show that, for the complete Pauli measurement family, the upper bound can be sharpened by considering arbitrary quantum preparations.

\paragraph{Quantum benchmark.}
Let $P_1,\ldots,P_D$ denote all non-identity $m$-qubit Hermitian Pauli observables. We define the optimal quantum bias for this fixed measurement family as
\begin{equation}
    \beta^Q_{\mathrm{Pauli}}(m,D)
    =
    \max_{\{\rho_x\}}
    \min_{\substack{x\in\{0,1\}^D\\y\in\{1,\ldots,D\}}}
    \left[
    (-1)^{x_y}\operatorname{Tr}(\rho_xP_y)
    \right],
\end{equation}
where the maximization is over arbitrary $m$-qubit quantum preparations. An explicit construction of a maximal QRAC using the complete set of nonidentity Pauli observables was presented in Ref.~\cite{Liabotro_2017}, yielding the success probability given in Eq.~(34) of that work. We use this construction as a lower bound on the general quantum value:
\begin{equation}
    \beta^Q_{\mathrm{Pauli}}(m,D) \geq\frac{1}{\sqrt{D(d-1)}} =\frac{1}{(d-1)\sqrt{d+1}}.
    \label{eq:quantum_pauli_lower}
\end{equation}
In the same reference, the author derived an analytical upper bound for maximal QRACs using the complete set of non-identity Pauli observables. However, more recently, general minimax characterizations of worst-case QRACs have been developed in Ref.~\cite{akibue2026}. In what follows, we use the latter characterizations to obtain a complementary analytical upper bound for the complete Pauli measurement family RAC.  Our approach exploits the average absolute expectation value of Pauli observables over Haar-random pure states, yielding a bound with a tighter asymptotic scaling than the previously established bound of Ref.~\cite{Liabotro_2017}, $\beta_{\mathrm{Liab}}^Q =  [(\sqrt{3} + 1)^m - 1]^{-1}$. Interestingly, it was conjectured in the same work that the proposed bound was attainable.  In particular, our bound becomes incompatible with the conjectured value for $m\geq 7$, thereby independently disproving its attainability in this regime for Pauli measurements. We note that an explicit counterexample for $m=3$, based on a machine-checked formalization, has also been established in Ref.~\cite{trureturingPauliRefutation}.

For a fixed set of observables $\{B_y\}$, the optimal quantum bias can be written as (see Eq.~(12) of Ref.~\cite{akibue2026})
\begin{equation}
\beta_{\{B_y\}}^Q(m,D)=\min_{x\in\{0,1\}^D}\min_{q\in\Delta_D}\lambda_{\max}\left(\sum_{y=1}^D(-1)^{x_y}q_yB_y\right),\label{eq:gen_minmax}
\end{equation}
where $\Delta_D$ denotes the probability simplex in $D$ dimensions and $\lambda_{max}(C)$ denotes the largest eigenvalue of $C$. To obtain an upper bound for our case of interest, consider an arbitrary pure state $\ket{\psi}$. Now define
\begin{equation}
    r_y= \bra{\psi}P_y\ket{\psi}\quad \text{and}\quad L_\psi=\sum_{y=1}^{D}\left|r_y\right|. 
\end{equation}
Expanding $\ket{\psi}$ in the Pauli basis gives
\begin{equation} 
\sum_{y=1}^D r_yP_y = d\ket{\psi}\bra{\psi}-\mathbb I.
\end{equation}
Now make the specific choice (for $r_y=0$, the choice is irrelevant)
\begin{equation}
(-1)^{x_y}=-\operatorname{sgn}(r_y),\qquad q_y=\frac{|r_y|}{L_\psi}.
\end{equation}
Note that $q=(q_1,\dots,q_D)\in \Delta_D$. Then, for these fixed choices of $x$ and $q$, and considering $B_y =P_y$, using Eq.~\eqref{eq:gen_minmax} gives
\begin{align}
    \beta_{\mathrm{Pauli}}^Q(m,D)&\leq\lambda_{\max}\left(\sum_{y=1}^D(-1)^{x_y}q_yP_y\right)\nonumber\\
    &=\frac{1}{L_\psi}\lambda_{\max} \left(\mathbb I- d\ket{\psi}\bra{\psi}\right) =\frac{1}{L_\psi}.
\end{align}
Since this holds for every pure state $\ket{\psi}$, it also holds for a state with $L_\psi$ at least as large as its Haar average, so $\beta^Q_{\mathrm{Pauli}}(m,D)\leq 1/\mathbb E_\psi(L_\psi)$.

Using the Haar-random Pauli-spectrum distribution shown in Eq.~(9) of  Ref.~\cite{Turkeshi_2025}, the average absolute expectation value of any non-identity Pauli operator is

\begin{align}
\mu_d&=\mathbb E_\psi\left[|\langle\psi|P|\psi\rangle|\right]\nonumber\\
&=\frac{\Gamma\left(\frac{d+1}{2}\right)}{\sqrt{\pi}\Gamma\left(\frac d2\right)}\int_{-1}^{1}|x|(1-x^2)^{\frac d2-1}\,dx=\frac{2\Gamma\left(\frac{d+1}{2}\right)}{d\sqrt{\pi}\Gamma\left(\frac d2\right)}.
\end{align}
resulting in that $\mathbb E_{\psi}(L_\psi)=D\mu_d$, which gives the upper bound, 
\begin{equation}
    \beta_{\mathrm{Pauli}}^Q(m,D)\leq\frac{1}{D\mu_d}.\label{eq:quantum_pauli_upper}
\end{equation}
Therefore, in the quantum case restricted to Pauli measurements we have that 
\begin{equation}
    \frac{1}{(d-1)\sqrt{d+1}}\leq\beta_{\mathrm{Pauli}}^Q(m,D)\leq\frac{1}{D\mu_d}\label{eq:sandwichQpauli}
\end{equation}

We now consider the asymptotic limit. For large $m$, using the expansion $\Gamma(z+a)/\Gamma(z+b) \approx z^{a-b}$~\cite{Erdelyi_1951}, one gets that 
\begin{equation}
    \mu_d\approx\sqrt{\frac{2}{\pi d}}.
\end{equation}
With this, in the large $m$ approximation, Eq.~\eqref{eq:sandwichQpauli} reads
\begin{equation}
    \frac{1}{d^{3/2}} \leq \beta^Q_{\mathrm{Pauli}}(m,D)\leq\sqrt{\frac{\pi}{2}}\frac{1}{d^{3/2}}.    \label{eq:quantum_pauli_bounds}
\end{equation}
Therefore, the optimal quantum bias for the complete Pauli measurement family scales as
$\beta^Q_{\mathrm{Pauli}}(m,D)=O(d^{-3/2})$.
We can now compare the optimal quantum and stabilizer values. Since the stabilizer polytope is a subset of the set of all quantum states, we necessarily have $\beta_{\mathrm{Pauli}}^{\mathrm{STABP}}(m,D)\leq\beta^Q_{\mathrm{Pauli}}(m,D)$. Combining the quantum upper bound with the inradius-based stabilizer lower bound in Eq.~\eqref{eq:scaling_stab_lower}, we obtain
\begin{equation}
    1 \leq \frac{\beta^Q_{\mathrm{Pauli}}(m,D)} {\beta_{\mathrm{Pauli}}^{\mathrm{STABP}}(m,D)} \leq \frac{\sqrt{d-a_\ast}} {\sqrt{a_\ast }D\,\mu_d}.
\end{equation}
In the large-$m$ limit, this gives
\begin{equation}
     \frac{\beta^Q_{\mathrm{Pauli}}(m,D)} {\beta_{\mathrm{Pauli}}^{\mathrm{STABP}}(m,D)} \leq \sqrt{\frac{\pi}{2a_\ast}} \simeq 1.85. \label{eq:constant_separation}
\end{equation}
Thus, for the complete Pauli measurement family, arbitrary quantum preparations and mixed stabilizer preparations achieve worst-case biases that are within a constant factor of one another. In particular, both values scale as $d^{-3/2}=2^{-3m/2}$. This means that, when Bob is restricted to Pauli measurements, magic in the preparations cannot provide more than a constant-factor improvement over mixtures of stabilizer states in the worst-case bias. Whether this remains true for non-Pauli measurements is an open question, and our results suggest that measurements beyond the stabilizer formalism should be considered to obtain a larger separation.

Moreover, the general quantum upper bound provides a tighter upper bound on the stabilizer value in this full-Pauli case than the general estimate in Eq.~\eqref{eq:scaling_stab_Bound}. The bounds obtained here do not determine the exact optimal values, but they establish their asymptotic scaling and constrain their ratio by a constant independent of the system size.

\section{Discussion}\label{sec:discussion}
In this work, we proposed the use of worst-case probabilities in random access codes as a tool to certify non-stabilizerness in many-qubit systems. We introduced a simple condition, the stabilizer pigeonhole criterion, which exploits the finite number of pure stabilizer states and of their marginals. With it, we showed how to certify $m$-qubit $W$-states ($m\geq3$) using only three single-qubit measurements, under the assumptions of the system dimension and of the specific Clifford encodings and measurements.

We then asked whether magic can be certified in global states whose marginals are only guaranteed to belong to the stabilizer polytope. We showed that this is possible: by letting Bob's measurements act on different subsets of the received qubits depending on the input $y$, a single experiment certifies magic in the global state even though every $k$-qubit marginal is a stabilizer state, a feature that cannot be detected from any single marginal. While this setup reveals the global magic, it depends on pairwise anti-commuting Paulis with distinct supports, which are not guaranteed to exist for arbitrary $(m,n)$.

Therefore, our next step was to study worst-case random access codes with $n$ arbitrary Pauli measurements. In this scenario, we derived upper and lower bounds for the stabilizer value: the maximum worst-case bias attainable with mixtures of $m$-qubit stabilizer states and $n$ Pauli measurements. For $n=4^m-1$, both bounds scale as $2^{-3m/2}$ and differ only by a factor linear in $m$, so the upper bound is tight up to this factor. We also derived bounds for general quantum states restricted to the same measurements, which allows us to compare stabilizer and quantum preparations on equal footing. We found that, for large $m$, the ratio between the quantum and the mixed-stabilizer bias is at most $\approx1.85$. Operationally, this means that by considering the full set of Pauli measurements, magic in the preparations yields at most a constant-factor advantage over stabilizer mixtures. Whether non-Pauli measurements are needed to obtain a larger separation requires further investigation.

Overall, our results show how to certify magic in $m$-qubit states with mild assumptions on the state preparations. We note, however, that for some of the protocols we assume specific structure in the measurements. An interesting problem for future research is to derive tight bounds for general measurements, moving as close as possible to fully device-independent magic certification.

Additionally, it remains an open task to close the gap between the upper and lower bounds obtained in this work, which would yield the exact stabilizer value for random access codes restricted to Pauli measurements.  Another relevant problem is to fully characterize the set of prepare-and-measure  correlations generated by global $m$-qubit quantum states whose every $k$-qubit marginal belongs to the stabilizer polytope. Solving this would be a significant contribution to the study of genuinely multipartite magic in its own right.
\\
\\
\acknowledgements
The author thanks Rafael Chaves for insightful discussions and support that motivated this research. S.Z. acknowledges the support of the Natural Sciences and Engineering Research Council of Canada (NSERC), (CIRTA-P-614385-2026). \textit{AI statement}: ChatGPT and Gemini were used to assist with derivations and manuscript preparation. All results were independently verified by the author, who takes full responsibility for the final content of this work.

\appendix
\section{Proof of Lemma~\ref{lemma:pauli_norm}}\label{app:pauli_norm}
Write $\rho =\ket{\psi}\bra{\psi} = \frac{1}{2^m}\sum_{g\in G}g$, where $G$ is the stabilizer group of $\ket{\psi}$.
Then
\begin{equation}
   |\bra{\psi}P\ket{\psi}|^2 =  |\Tr(\rho P)|^2 = \frac{1}{4^m}|\sum_{g\in G}\Tr(gP)|^2,
\end{equation}
for $P$ a Pauli operator. Then, if $\pm P\in G$, $\sum_{g\in G}\Tr(gP)=\pm2^m$, otherwise $\sum_{g\in G}\Tr(gP)=0$. We can write this compactly by introducing the indicator function $\mathds{1}_{\pm P\in G}$. Furthermore, since the only possible expected values of P over stabilizer states are $\{0,1,-1\}$, then
\begin{equation} 
    |\bra{\psi}P\ket{\psi}|^2 = |\bra{\psi}P\ket{\psi}|  = \mathds{1}_{\pm P\in G}
\end{equation}

One notes that the stabilizer group contains exactly $d=2^m$ Pauli elements, including the identity (each Pauli operator appears at most once up to sign, since $-\mathbb I\notin G$). Therefore summing over all Paulis minus the identity gives
\begin{equation}
    \sum_{P\neq\mathbb I} |\bra{\psi}P\ket{\psi}| = d-1.
\end{equation}

\section{Proof of upper bound~\eqref{eq:upperbound2}}\label{app:upper_bound}
In this appendix we give an explicit derivation of the bound given in Eq.~\eqref{eq:upperbound2}.
We begin by  summing over $y$ in both sides of Eq.~\eqref{eq:useful_bound}, obtaining 
\begin{equation}
    n\beta(m,n) \leq \sum_{y=1}^n(-1)^{x_y}\Tr(\rho_xP_y)\;\leq\;\max_{\phi\in\mathrm{STAB}_m}\sum_{y=1}^n(-1)^{x_y}\bra{\phi}P_y\ket{\phi}.\label{eq:beta_2ndbound_app}
\end{equation}
Now, assume that we choose a bit-string $x$ uniformly at random. Then we can define a random variable over this bitstring
\begin{equation}
    X_\phi(x) =\sum_{y=1}^n(-1)^{x_y}\bra{\phi}P_y\ket{\phi}=\sum_{y=1}^n\alpha_y,
\end{equation}
where $\alpha_y$ are $n$ independent random variables. Each of these random variables has a zero expectation value
\begin{equation}
    \mathbb E_x(\alpha_y) = \mathbb E_x( (-1)^{x_y})\bra{\phi}P_y\ket{\phi} = 0.
\end{equation}
Furthermore, $\alpha_y$ is bounded in $[a_y,b_y]\in[-|\bra{\phi}P_y\ket{\phi}|,|\bra{\phi}P_y\ket{\phi}|]$ so $b_y-a_y=2|\bra{\phi}P_y\ket{\phi}|$. Therefore, we can apply Hoeffding's Lemma~\cite[p.~21]{massart2007} to each of these variables giving 
\begin{equation}
    \mathbb E_x(\exp{\lambda\alpha_y})\leq \exp{\frac{\lambda^2|\bra{\phi}P_y\ket{\phi}|^2}{2}},\quad\forall  \lambda\in \mathbb R.
\end{equation}
This implies
\begin{align}
    \mathbb E_x\left(\exp{\lambda X_\phi(x)} \right)&= \prod_{y=1}^n \mathbb E_x(\exp{\lambda\alpha_y}) \nonumber \\
    &\leq \prod_{y=1}^n \exp{\frac{\lambda^2|\bra{\phi}P_y\ket{\phi}|^2}{2}},\nonumber \\
    &\leq\exp{\frac{\lambda^2}{2}\sum_{y=1}^n|\bra{\phi}P_y\ket{\phi}|^2}\nonumber\\
    &\leq\exp{\frac{\lambda^2(d-1)}{2}}.\label{eq:Hoeffdings_bound}
\end{align}
The last step uses $|\bra{\phi}P_y\ket{\phi}|^2\le|\bra{\phi}P_y\ket{\phi}|$, the $P_y$ being distinct, and Lemma~\ref{lemma:pauli_norm}. Now we restrict to $\lambda>0$ so that
\begin{align}
    \exp{\lambda \max_\phi X_\phi(x)}  &= \max_\phi\exp{\lambda  X_\phi(x)}\nonumber\\&\leq\sum_{\phi=1}^{N_m}\exp{\lambda  X_\phi(x)},
\end{align}
since the sum contains the maximum. Then
\begin{equation}
    \mathbb E_x(\exp{\lambda \max_\phi X_\phi(x)})\leq\mathbb E_x (\sum_{\phi=1}^{N_m}\exp{\lambda  X_\phi(x)}).
\end{equation}
Applying Jensen's inequality to the left side, and Hoeffding's bound~\eqref{eq:Hoeffdings_bound} to the right side, we get
\begin{equation}
    \exp{\lambda\mathbb E_x(\max_\phi X_\phi(x))}\leq N_m \exp{\frac{\lambda^2(d-1)}{2}}.
\end{equation}
By taking the logarithm, dividing by $\lambda>0$, we get
\begin{equation}
    \mathbb E_x(\max_\phi X_\phi(x))\leq\frac{\ln N_m}{\lambda} + \frac{\lambda(d-1)}{2}.
\end{equation}
Finally, we can optimize the right hand side by taking the derivative with respect to $\lambda$ and equating to zero. Solving for $\lambda$ we find that $\lambda = \sqrt{2\ln (N_m)/(d-1)}$. This value gives the optimized upper bound
\begin{equation}
    \mathbb E_x (\max_\phi X_\phi(x))\leq\sqrt{2\ln N_m(d-1)}.
\end{equation}
Finally, using Eq.~\eqref{eq:beta_2ndbound_app}, valid for every $x$, hence also for the average over $x$ (more precisely $n\beta\le\min_x\max_\phi X_\phi(x)\le\mathbb{E}_x\max_\phi X_\phi(x)$), we get
\begin{equation}
    n\beta^{\mathrm{STABP}}_{\mathrm{Pauli}}(m,n)\leq\sqrt{2(d-1)\ln N_m}.
\end{equation}

\bibliography{2-references}

\begin{thebibliography}{34}%
\makeatletter
\providecommand \@ifxundefined [1]{%
 \@ifx{#1\undefined}
}%
\providecommand \@ifnum [1]{%
 \ifnum #1\expandafter \@firstoftwo
 \else \expandafter \@secondoftwo
 \fi
}%
\providecommand \@ifx [1]{%
 \ifx #1\expandafter \@firstoftwo
 \else \expandafter \@secondoftwo
 \fi
}%
\providecommand \natexlab [1]{#1}%
\providecommand \enquote  [1]{``#1''}%
\providecommand \bibnamefont  [1]{#1}%
\providecommand \bibfnamefont [1]{#1}%
\providecommand \citenamefont [1]{#1}%
\providecommand \href@noop [0]{\@secondoftwo}%
\providecommand \href [0]{\begingroup \@sanitize@url \@href}%
\providecommand \@href[1]{\@@startlink{#1}\@@href}%
\providecommand \@@href[1]{\endgroup#1\@@endlink}%
\providecommand \@sanitize@url [0]{\catcode `\\12\catcode `\$12\catcode `\&12\catcode `\#12\catcode `\^12\catcode `\_12\catcode `\%12\relax}%
\providecommand \@@startlink[1]{}%
\providecommand \@@endlink[0]{}%
\providecommand \url  [0]{\begingroup\@sanitize@url \@url }%
\providecommand \@url [1]{\endgroup\@href {#1}{\urlprefix }}%
\providecommand \urlprefix  [0]{URL }%
\providecommand \Eprint [0]{\href }%
\providecommand \doibase [0]{https://doi.org/}%
\providecommand \selectlanguage [0]{\@gobble}%
\providecommand \bibinfo  [0]{\@secondoftwo}%
\providecommand \bibfield  [0]{\@secondoftwo}%
\providecommand \translation [1]{[#1]}%
\providecommand \BibitemOpen [0]{}%
\providecommand \bibitemStop [0]{}%
\providecommand \bibitemNoStop [0]{.\EOS\space}%
\providecommand \EOS [0]{\spacefactor3000\relax}%
\providecommand \BibitemShut  [1]{\csname bibitem#1\endcsname}%
\let\auto@bib@innerbib\@empty
\bibitem [{\citenamefont {Bravyi}\ and\ \citenamefont {Kitaev}(2005)}]{Bravyi_2005}%
  \BibitemOpen
  \bibfield  {author} {\bibinfo {author} {\bibfnamefont {S.}~\bibnamefont {Bravyi}}\ and\ \bibinfo {author} {\bibfnamefont {A.}~\bibnamefont {Kitaev}},\ }\bibfield  {title} {\bibinfo {title} {Universal quantum computation with ideal clifford gates and noisy ancillas},\ }\href {https://doi.org/10.1103/PhysRevA.71.022316} {\bibfield  {journal} {\bibinfo  {journal} {Phys. Rev. A}\ }\textbf {\bibinfo {volume} {71}},\ \bibinfo {pages} {022316} (\bibinfo {year} {2005})}\BibitemShut {NoStop}%
\bibitem [{\citenamefont {Veitch}\ \emph {et~al.}(2014)\citenamefont {Veitch}, \citenamefont {Mousavian}, \citenamefont {Gottesman},\ and\ \citenamefont {Emerson}}]{Veitch_2014}%
  \BibitemOpen
  \bibfield  {author} {\bibinfo {author} {\bibfnamefont {V.}~\bibnamefont {Veitch}}, \bibinfo {author} {\bibfnamefont {S.~A.~H.}\ \bibnamefont {Mousavian}}, \bibinfo {author} {\bibfnamefont {D.}~\bibnamefont {Gottesman}},\ and\ \bibinfo {author} {\bibfnamefont {J.}~\bibnamefont {Emerson}},\ }\bibfield  {title} {\bibinfo {title} {The resource theory of stabilizer quantum computation},\ }\href {https://doi.org/10.1088/1367-2630/16/1/013009} {\bibfield  {journal} {\bibinfo  {journal} {New Journal of Physics}\ }\textbf {\bibinfo {volume} {16}},\ \bibinfo {pages} {013009} (\bibinfo {year} {2014})}\BibitemShut {NoStop}%
\bibitem [{\citenamefont {Gottesman}(1997)}]{gottesman1997}%
  \BibitemOpen
  \bibfield  {author} {\bibinfo {author} {\bibfnamefont {D.}~\bibnamefont {Gottesman}},\ }\href@noop {} {\emph {\bibinfo {title} {Stabilizer codes and quantum error correction}}}\ (\bibinfo  {publisher} {California Institute of Technology},\ \bibinfo {year} {1997})\BibitemShut {NoStop}%
\bibitem [{\citenamefont {Gottesman}(1998)}]{gottesman1998}%
  \BibitemOpen
  \bibfield  {author} {\bibinfo {author} {\bibfnamefont {D.}~\bibnamefont {Gottesman}},\ }\bibfield  {title} {\bibinfo {title} {The heisenberg representation of quantum computers},\ }\href@noop {} {\bibfield  {journal} {\bibinfo  {journal} {arXiv preprint quant-ph/9807006}\ } (\bibinfo {year} {1998})}\BibitemShut {NoStop}%
\bibitem [{\citenamefont {Howard}\ \emph {et~al.}(2014)\citenamefont {Howard}, \citenamefont {Wallman}, \citenamefont {Veitch},\ and\ \citenamefont {Emerson}}]{Howard2014}%
  \BibitemOpen
  \bibfield  {author} {\bibinfo {author} {\bibfnamefont {M.}~\bibnamefont {Howard}}, \bibinfo {author} {\bibfnamefont {J.}~\bibnamefont {Wallman}}, \bibinfo {author} {\bibfnamefont {V.}~\bibnamefont {Veitch}},\ and\ \bibinfo {author} {\bibfnamefont {J.}~\bibnamefont {Emerson}},\ }\bibfield  {title} {\bibinfo {title} {Contextuality supplies the `magic' for quantum computation},\ }\href {https://doi.org/10.1038/nature13460} {\bibfield  {journal} {\bibinfo  {journal} {Nature}\ }\textbf {\bibinfo {volume} {510}},\ \bibinfo {pages} {351} (\bibinfo {year} {2014})}\BibitemShut {NoStop}%
\bibitem [{\citenamefont {Leone}\ \emph {et~al.}(2022)\citenamefont {Leone}, \citenamefont {Oliviero},\ and\ \citenamefont {Hamma}}]{Leone2022}%
  \BibitemOpen
  \bibfield  {author} {\bibinfo {author} {\bibfnamefont {L.}~\bibnamefont {Leone}}, \bibinfo {author} {\bibfnamefont {S.~F.~E.}\ \bibnamefont {Oliviero}},\ and\ \bibinfo {author} {\bibfnamefont {A.}~\bibnamefont {Hamma}},\ }\bibfield  {title} {\bibinfo {title} {Stabilizer r\'enyi entropy},\ }\href {https://doi.org/10.1103/PhysRevLett.128.050402} {\bibfield  {journal} {\bibinfo  {journal} {Phys. Rev. Lett.}\ }\textbf {\bibinfo {volume} {128}},\ \bibinfo {pages} {050402} (\bibinfo {year} {2022})}\BibitemShut {NoStop}%
\bibitem [{\citenamefont {Oliviero}\ \emph {et~al.}(2022)\citenamefont {Oliviero}, \citenamefont {Leone},\ and\ \citenamefont {Hamma}}]{Oliveiro_2022}%
  \BibitemOpen
  \bibfield  {author} {\bibinfo {author} {\bibfnamefont {S.~F.~E.}\ \bibnamefont {Oliviero}}, \bibinfo {author} {\bibfnamefont {L.}~\bibnamefont {Leone}},\ and\ \bibinfo {author} {\bibfnamefont {A.}~\bibnamefont {Hamma}},\ }\bibfield  {title} {\bibinfo {title} {Magic-state resource theory for the ground state of the transverse-field ising model},\ }\href {https://doi.org/10.1103/PhysRevA.106.042426} {\bibfield  {journal} {\bibinfo  {journal} {Phys. Rev. A}\ }\textbf {\bibinfo {volume} {106}},\ \bibinfo {pages} {042426} (\bibinfo {year} {2022})}\BibitemShut {NoStop}%
\bibitem [{\citenamefont {Liu}\ and\ \citenamefont {Winter}(2022)}]{Liu_2022}%
  \BibitemOpen
  \bibfield  {author} {\bibinfo {author} {\bibfnamefont {Z.-W.}\ \bibnamefont {Liu}}\ and\ \bibinfo {author} {\bibfnamefont {A.}~\bibnamefont {Winter}},\ }\bibfield  {title} {\bibinfo {title} {Many-body quantum magic},\ }\href {https://doi.org/10.1103/PRXQuantum.3.020333} {\bibfield  {journal} {\bibinfo  {journal} {PRX Quantum}\ }\textbf {\bibinfo {volume} {3}},\ \bibinfo {pages} {020333} (\bibinfo {year} {2022})}\BibitemShut {NoStop}%
\bibitem [{\citenamefont {Mac\^edo}\ \emph {et~al.}(2025)\citenamefont {Mac\^edo}, \citenamefont {Andriolo}, \citenamefont {Zamora}, \citenamefont {Poderini},\ and\ \citenamefont {Chaves}}]{Macedo_2025}%
  \BibitemOpen
  \bibfield  {author} {\bibinfo {author} {\bibfnamefont {R.~A.}\ \bibnamefont {Mac\^edo}}, \bibinfo {author} {\bibfnamefont {P.}~\bibnamefont {Andriolo}}, \bibinfo {author} {\bibfnamefont {S.}~\bibnamefont {Zamora}}, \bibinfo {author} {\bibfnamefont {D.}~\bibnamefont {Poderini}},\ and\ \bibinfo {author} {\bibfnamefont {R.}~\bibnamefont {Chaves}},\ }\bibfield  {title} {\bibinfo {title} {Witnessing nonstabilizerness with bell inequalities},\ }\href {https://doi.org/10.1103/6srg-723m} {\bibfield  {journal} {\bibinfo  {journal} {Phys. Rev. A}\ }\textbf {\bibinfo {volume} {112}},\ \bibinfo {pages} {L050401} (\bibinfo {year} {2025})}\BibitemShut {NoStop}%
\bibitem [{\citenamefont {Zamora}\ \emph {et~al.}(2025)\citenamefont {Zamora}, \citenamefont {Mac{\^e}do}, \citenamefont {Sarubi}, \citenamefont {Alves}, \citenamefont {Poderini},\ and\ \citenamefont {Chaves}}]{Zamora2025}%
  \BibitemOpen
  \bibfield  {author} {\bibinfo {author} {\bibfnamefont {S.}~\bibnamefont {Zamora}}, \bibinfo {author} {\bibfnamefont {R.~A.}\ \bibnamefont {Mac{\^e}do}}, \bibinfo {author} {\bibfnamefont {T.~S.}\ \bibnamefont {Sarubi}}, \bibinfo {author} {\bibfnamefont {M.}~\bibnamefont {Alves}}, \bibinfo {author} {\bibfnamefont {D.}~\bibnamefont {Poderini}},\ and\ \bibinfo {author} {\bibfnamefont {R.}~\bibnamefont {Chaves}},\ }\bibfield  {title} {\bibinfo {title} {Semi-device-independent nonstabilizerness certification in the prepare-and-measure scenario},\ }\href {https://doi.org/10.1103/wm7m-xnfq} {\bibfield  {journal} {\bibinfo  {journal} {Phys. Rev. A}\ }\textbf {\bibinfo {volume} {112}},\ \bibinfo {pages} {042410} (\bibinfo {year} {2025})}\BibitemShut {NoStop}%
\bibitem [{\citenamefont {Stratton}\ \emph {et~al.}(2026)\citenamefont {Stratton}, \citenamefont {Zamora},\ and\ \citenamefont {Chaves}}]{stratton2026}%
  \BibitemOpen
  \bibfield  {author} {\bibinfo {author} {\bibfnamefont {B.}~\bibnamefont {Stratton}}, \bibinfo {author} {\bibfnamefont {S.}~\bibnamefont {Zamora}},\ and\ \bibinfo {author} {\bibfnamefont {R.}~\bibnamefont {Chaves}},\ }\href {https://arxiv.org/abs/2607.27339} {\bibinfo {title} {Quantum state discrimination with stabilizer circuits}} (\bibinfo {year} {2026}),\ \Eprint {https://arxiv.org/abs/2607.27339} {arXiv:2607.27339 [quant-ph]} \BibitemShut {NoStop}%
\bibitem [{\citenamefont {Zhu}\ \emph {et~al.}(2024)\citenamefont {Zhu}, \citenamefont {Liu}, \citenamefont {Zhu},\ and\ \citenamefont {Wang}}]{Zhu_2024}%
  \BibitemOpen
  \bibfield  {author} {\bibinfo {author} {\bibfnamefont {C.}~\bibnamefont {Zhu}}, \bibinfo {author} {\bibfnamefont {Z.}~\bibnamefont {Liu}}, \bibinfo {author} {\bibfnamefont {C.}~\bibnamefont {Zhu}},\ and\ \bibinfo {author} {\bibfnamefont {X.}~\bibnamefont {Wang}},\ }\bibfield  {title} {\bibinfo {title} {Limitations of classically simulable measurements for quantum state discrimination},\ }\href {https://doi.org/10.1103/PhysRevLett.133.010202} {\bibfield  {journal} {\bibinfo  {journal} {Phys. Rev. Lett.}\ }\textbf {\bibinfo {volume} {133}},\ \bibinfo {pages} {010202} (\bibinfo {year} {2024})}\BibitemShut {NoStop}%
\bibitem [{\citenamefont {Chowdhury}\ \emph {et~al.}(2026)\citenamefont {Chowdhury}, \citenamefont {Naik}, \citenamefont {Chakraborty}, \citenamefont {Patra}, \citenamefont {Ghosh}, \citenamefont {Ghosal}, \citenamefont {Banik},\ and\ \citenamefont {Maity}}]{Chowdhury_2026}%
  \BibitemOpen
  \bibfield  {author} {\bibinfo {author} {\bibfnamefont {S.~R.}\ \bibnamefont {Chowdhury}}, \bibinfo {author} {\bibfnamefont {S.~G.}\ \bibnamefont {Naik}}, \bibinfo {author} {\bibfnamefont {A.}~\bibnamefont {Chakraborty}}, \bibinfo {author} {\bibfnamefont {R.~K.}\ \bibnamefont {Patra}}, \bibinfo {author} {\bibfnamefont {S.~B.}\ \bibnamefont {Ghosh}}, \bibinfo {author} {\bibfnamefont {P.}~\bibnamefont {Ghosal}}, \bibinfo {author} {\bibfnamefont {M.}~\bibnamefont {Banik}},\ and\ \bibinfo {author} {\bibfnamefont {A.~G.}\ \bibnamefont {Maity}},\ }\bibfield  {title} {\bibinfo {title} {Gottesman-knill limit on one-way communication complexity: Tracing the quantum advantage down to magic resources},\ }\href {https://doi.org/10.1103/x741-4kd1} {\bibfield  {journal} {\bibinfo  {journal} {Phys. Rev. A}\ }\textbf {\bibinfo {volume} {113}},\ \bibinfo {pages} {042607} (\bibinfo {year} {2026})}\BibitemShut {NoStop}%
\bibitem [{\citenamefont {Girish}\ \emph {et~al.}(2026)\citenamefont {Girish}, \citenamefont {May}, \citenamefont {Parham},\ and\ \citenamefont {Yuen}}]{Girish_2025}%
  \BibitemOpen
  \bibfield  {author} {\bibinfo {author} {\bibfnamefont {U.}~\bibnamefont {Girish}}, \bibinfo {author} {\bibfnamefont {A.}~\bibnamefont {May}}, \bibinfo {author} {\bibfnamefont {N.}~\bibnamefont {Parham}},\ and\ \bibinfo {author} {\bibfnamefont {H.}~\bibnamefont {Yuen}},\ }\bibfield  {title} {\bibinfo {title} {Magic and communication complexity},\ }in\ \href {https://doi.org/10.1145/3798129.3800799} {\emph {\bibinfo {booktitle} {Proceedings of the 58th Annual ACM Symposium on Theory of Computing}}},\ \bibinfo {series and number} {STOC '26}\ (\bibinfo  {publisher} {Association for Computing Machinery},\ \bibinfo {address} {New York, NY, USA},\ \bibinfo {year} {2026})\ p.\ \bibinfo {pages} {845–856}\BibitemShut {NoStop}%
\bibitem [{\citenamefont {Vairogs}\ and\ \citenamefont {Yan}(2025)}]{Vairogs_2025}%
  \BibitemOpen
  \bibfield  {author} {\bibinfo {author} {\bibfnamefont {C.}~\bibnamefont {Vairogs}}\ and\ \bibinfo {author} {\bibfnamefont {B.}~\bibnamefont {Yan}},\ }\bibfield  {title} {\bibinfo {title} {Extracting randomness from magic quantum states},\ }\href {https://doi.org/10.1103/3ttm-vhdt} {\bibfield  {journal} {\bibinfo  {journal} {Phys. Rev. Res.}\ }\textbf {\bibinfo {volume} {7}},\ \bibinfo {pages} {L022069} (\bibinfo {year} {2025})}\BibitemShut {NoStop}%
\bibitem [{\citenamefont {Varela}\ \emph {et~al.}(2026)\citenamefont {Varela}, \citenamefont {Keller}, \citenamefont {Junior}, \citenamefont {Moreira}, \citenamefont {Chaves},\ and\ \citenamefont {Mac{\^e}do}}]{Varela_2026}%
  \BibitemOpen
  \bibfield  {author} {\bibinfo {author} {\bibfnamefont {J.~M.}\ \bibnamefont {Varela}}, \bibinfo {author} {\bibfnamefont {L.~L.}\ \bibnamefont {Keller}}, \bibinfo {author} {\bibfnamefont {A.~d.~O.}\ \bibnamefont {Junior}}, \bibinfo {author} {\bibfnamefont {D.~A.}\ \bibnamefont {Moreira}}, \bibinfo {author} {\bibfnamefont {R.}~\bibnamefont {Chaves}},\ and\ \bibinfo {author} {\bibfnamefont {R.~A.}\ \bibnamefont {Mac{\^e}do}},\ }\bibfield  {title} {\bibinfo {title} {Predicting magic from very few measurements},\ }\href {https://doi.org/10.1103/2xnf-jcyn} {\bibfield  {journal} {\bibinfo  {journal} {PRX Quantum}\ }\textbf {\bibinfo {volume} {7}},\ \bibinfo {pages} {043009} (\bibinfo {year} {2026})}\BibitemShut {NoStop}%
\bibitem [{\citenamefont {Bonet-Monroig}\ \emph {et~al.}(2020)\citenamefont {Bonet-Monroig}, \citenamefont {Babbush},\ and\ \citenamefont {O'Brien}}]{Bonet-Monroig_2020}%
  \BibitemOpen
  \bibfield  {author} {\bibinfo {author} {\bibfnamefont {X.}~\bibnamefont {Bonet-Monroig}}, \bibinfo {author} {\bibfnamefont {R.}~\bibnamefont {Babbush}},\ and\ \bibinfo {author} {\bibfnamefont {T.~E.}\ \bibnamefont {O'Brien}},\ }\bibfield  {title} {\bibinfo {title} {Nearly optimal measurement scheduling for partial tomography of quantum states},\ }\href {https://doi.org/10.1103/PhysRevX.10.031064} {\bibfield  {journal} {\bibinfo  {journal} {Phys. Rev. X}\ }\textbf {\bibinfo {volume} {10}},\ \bibinfo {pages} {031064} (\bibinfo {year} {2020})}\BibitemShut {NoStop}%
\bibitem [{\citenamefont {Liab{\o{}}tr{\o{}}}(2017)}]{Liabotro_2017}%
  \BibitemOpen
  \bibfield  {author} {\bibinfo {author} {\bibfnamefont {O.}~\bibnamefont {Liab{\o{}}tr{\o{}}}},\ }\bibfield  {title} {\bibinfo {title} {Improved classical and quantum random access codes},\ }\href {https://doi.org/10.1103/PhysRevA.95.052315} {\bibfield  {journal} {\bibinfo  {journal} {Phys. Rev. A}\ }\textbf {\bibinfo {volume} {95}},\ \bibinfo {pages} {052315} (\bibinfo {year} {2017})}\BibitemShut {NoStop}%
\bibitem [{\citenamefont {Ekert}\ \emph {et~al.}()\citenamefont {Ekert}, \citenamefont {Hosgood}, \citenamefont {Kay},\ and\ \citenamefont {Macchiavello}}]{qubitguide}%
  \BibitemOpen
  \bibfield  {author} {\bibinfo {author} {\bibfnamefont {A.}~\bibnamefont {Ekert}}, \bibinfo {author} {\bibfnamefont {T.}~\bibnamefont {Hosgood}}, \bibinfo {author} {\bibfnamefont {A.}~\bibnamefont {Kay}},\ and\ \bibinfo {author} {\bibfnamefont {C.}~\bibnamefont {Macchiavello}},\ }\href@noop {} {\bibinfo {title} {7.1 pauli groups}},\ \bibinfo {howpublished} {\url{https://qubit.guide/7.1-pauli-groups}}\BibitemShut {NoStop}%
\bibitem [{\citenamefont {Nielsen}\ and\ \citenamefont {Chuang}(2010)}]{NielsenChuand2010}%
  \BibitemOpen
  \bibfield  {author} {\bibinfo {author} {\bibfnamefont {M.~A.}\ \bibnamefont {Nielsen}}\ and\ \bibinfo {author} {\bibfnamefont {I.~L.}\ \bibnamefont {Chuang}},\ }\href@noop {} {\emph {\bibinfo {title} {Quantum Computation and Quantum Information: 10th Anniversary Edition}}}\ (\bibinfo  {publisher} {Cambridge University Press},\ \bibinfo {year} {2010})\BibitemShut {NoStop}%
\bibitem [{\citenamefont {de~Oliveira~Junior}\ \emph {et~al.}(2026)\citenamefont {de~Oliveira~Junior}, \citenamefont {Zamora}, \citenamefont {Brask}, \citenamefont {Chaves},\ and\ \citenamefont {Tavakoli}}]{junior2026}%
  \BibitemOpen
  \bibfield  {author} {\bibinfo {author} {\bibfnamefont {A.}~\bibnamefont {de~Oliveira~Junior}}, \bibinfo {author} {\bibfnamefont {S.}~\bibnamefont {Zamora}}, \bibinfo {author} {\bibfnamefont {J.~B.}\ \bibnamefont {Brask}}, \bibinfo {author} {\bibfnamefont {R.}~\bibnamefont {Chaves}},\ and\ \bibinfo {author} {\bibfnamefont {A.}~\bibnamefont {Tavakoli}},\ }\href {https://arxiv.org/abs/2610.20725} {\bibinfo {title} {Quantum prepare-and-measure scenarios: concepts and methods}} (\bibinfo {year} {2026}),\ \Eprint {https://arxiv.org/abs/2610.20725} {arXiv:2610.20725 [quant-ph]} \BibitemShut {NoStop}%
\bibitem [{\citenamefont {Tavakoli}\ \emph {et~al.}(2020)\citenamefont {Tavakoli}, \citenamefont {Zambrini~Cruzeiro}, \citenamefont {Bohr~Brask}, \citenamefont {Gisin},\ and\ \citenamefont {Brunner}}]{Tavakoli2020informationally}%
  \BibitemOpen
  \bibfield  {author} {\bibinfo {author} {\bibfnamefont {A.}~\bibnamefont {Tavakoli}}, \bibinfo {author} {\bibfnamefont {E.}~\bibnamefont {Zambrini~Cruzeiro}}, \bibinfo {author} {\bibfnamefont {J.}~\bibnamefont {Bohr~Brask}}, \bibinfo {author} {\bibfnamefont {N.}~\bibnamefont {Gisin}},\ and\ \bibinfo {author} {\bibfnamefont {N.}~\bibnamefont {Brunner}},\ }\bibfield  {title} {\bibinfo {title} {Informationally restricted quantum correlations},\ }\href {https://doi.org/10.22331/q-2020-09-24-332} {\bibfield  {journal} {\bibinfo  {journal} {{Quantum}}\ }\textbf {\bibinfo {volume} {4}},\ \bibinfo {pages} {332} (\bibinfo {year} {2020})}\BibitemShut {NoStop}%
\bibitem [{\citenamefont {Ambainis}\ \emph {et~al.}(2009)\citenamefont {Ambainis}, \citenamefont {Leung}, \citenamefont {Mancinska},\ and\ \citenamefont {Ozols}}]{ambainis2009}%
  \BibitemOpen
  \bibfield  {author} {\bibinfo {author} {\bibfnamefont {A.}~\bibnamefont {Ambainis}}, \bibinfo {author} {\bibfnamefont {D.}~\bibnamefont {Leung}}, \bibinfo {author} {\bibfnamefont {L.}~\bibnamefont {Mancinska}},\ and\ \bibinfo {author} {\bibfnamefont {M.}~\bibnamefont {Ozols}},\ }\href {https://arxiv.org/abs/0810.2937} {\bibinfo {title} {Quantum random access codes with shared randomness}} (\bibinfo {year} {2009}),\ \Eprint {https://arxiv.org/abs/0810.2937} {arXiv:0810.2937 [quant-ph]} \BibitemShut {NoStop}%
\bibitem [{\citenamefont {Iwama}\ \emph {et~al.}(2007)\citenamefont {Iwama}, \citenamefont {Nishimura}, \citenamefont {Raymond},\ and\ \citenamefont {Yamashita}}]{Iwama_2007}%
  \BibitemOpen
  \bibfield  {author} {\bibinfo {author} {\bibfnamefont {K.}~\bibnamefont {Iwama}}, \bibinfo {author} {\bibfnamefont {H.}~\bibnamefont {Nishimura}}, \bibinfo {author} {\bibfnamefont {R.}~\bibnamefont {Raymond}},\ and\ \bibinfo {author} {\bibfnamefont {S.}~\bibnamefont {Yamashita}},\ }\bibfield  {title} {\bibinfo {title} {Unbounded-error one-way classical and quantum communication complexity},\ }in\ \href@noop {} {\emph {\bibinfo {booktitle} {Automata, Languages and Programming}}},\ \bibinfo {editor} {edited by\ \bibinfo {editor} {\bibfnamefont {L.}~\bibnamefont {Arge}}, \bibinfo {editor} {\bibfnamefont {C.}~\bibnamefont {Cachin}}, \bibinfo {editor} {\bibfnamefont {T.}~\bibnamefont {Jurdzi{\'{n}}ski}},\ and\ \bibinfo {editor} {\bibfnamefont {A.}~\bibnamefont {Tarlecki}}}\ (\bibinfo  {publisher} {Springer Berlin Heidelberg},\ \bibinfo {address} {Berlin, Heidelberg},\ \bibinfo {year} {2007})\ pp.\ \bibinfo {pages} {110--121}\BibitemShut {NoStop}%
\bibitem [{\citenamefont {Audenaert}\ and\ \citenamefont {Plenio}(2005)}]{Audenaert_2005}%
  \BibitemOpen
  \bibfield  {author} {\bibinfo {author} {\bibfnamefont {K.~M.~R.}\ \bibnamefont {Audenaert}}\ and\ \bibinfo {author} {\bibfnamefont {M.~B.}\ \bibnamefont {Plenio}},\ }\bibfield  {title} {\bibinfo {title} {Entanglement on mixed stabilizer states: normal forms and reduction procedures},\ }\href {https://doi.org/10.1088/1367-2630/7/1/170} {\bibfield  {journal} {\bibinfo  {journal} {New Journal of Physics}\ }\textbf {\bibinfo {volume} {7}},\ \bibinfo {pages} {170} (\bibinfo {year} {2005})}\BibitemShut {NoStop}%
\bibitem [{\citenamefont {Gross}(2006)}]{Gross2006}%
  \BibitemOpen
  \bibfield  {author} {\bibinfo {author} {\bibfnamefont {D.}~\bibnamefont {Gross}},\ }\bibfield  {title} {\bibinfo {title} {Hudson's theorem for finite-dimensional quantum systems},\ }\href {https://doi.org/10.1063/1.2393152} {\bibfield  {journal} {\bibinfo  {journal} {Journal of Mathematical Physics}\ }\textbf {\bibinfo {volume} {47}},\ \bibinfo {pages} {122107} (\bibinfo {year} {2006})}\BibitemShut {NoStop}%
\bibitem [{\citenamefont {Singh}\ \emph {et~al.}(2023)\citenamefont {Singh}, \citenamefont {Chiang}, \citenamefont {Hsu}, \citenamefont {Ki}, \citenamefont {Goan},\ and\ \citenamefont {Hsieh}}]{Singh2023}%
  \BibitemOpen
  \bibfield  {author} {\bibinfo {author} {\bibfnamefont {A.~K.}\ \bibnamefont {Singh}}, \bibinfo {author} {\bibfnamefont {C.-F.}\ \bibnamefont {Chiang}}, \bibinfo {author} {\bibfnamefont {E.}~\bibnamefont {Hsu}}, \bibinfo {author} {\bibfnamefont {E.}~\bibnamefont {Ki}}, \bibinfo {author} {\bibfnamefont {H.-S.}\ \bibnamefont {Goan}},\ and\ \bibinfo {author} {\bibfnamefont {M.-H.}\ \bibnamefont {Hsieh}},\ }\bibfield  {title} {\bibinfo {title} {Counting stabilizer codes for arbitrary dimension},\ }\href {https://doi.org/10.22331/q-2023-07-06-1048} {\bibfield  {journal} {\bibinfo  {journal} {Quantum}\ }\textbf {\bibinfo {volume} {7}},\ \bibinfo {pages} {1048} (\bibinfo {year} {2023})}\BibitemShut {NoStop}%
\bibitem [{\citenamefont {Sarkar}\ and\ \citenamefont {van~den Berg}(2021)}]{Sarkar2021}%
  \BibitemOpen
  \bibfield  {author} {\bibinfo {author} {\bibfnamefont {R.}~\bibnamefont {Sarkar}}\ and\ \bibinfo {author} {\bibfnamefont {E.}~\bibnamefont {van~den Berg}},\ }\bibfield  {title} {\bibinfo {title} {On sets of maximally commuting and anticommuting pauli operators},\ }\href {https://doi.org/10.1007/s40687-020-00244-1} {\bibfield  {journal} {\bibinfo  {journal} {Research in the Mathematical Sciences}\ }\textbf {\bibinfo {volume} {8}},\ \bibinfo {pages} {14} (\bibinfo {year} {2021})}\BibitemShut {NoStop}%
\bibitem [{\citenamefont {Massart}(2007)}]{massart2007}%
  \BibitemOpen
  \bibfield  {author} {\bibinfo {author} {\bibfnamefont {P.}~\bibnamefont {Massart}},\ }\href@noop {} {\emph {\bibinfo {title} {Concentration Inequalities and Model Selection: Ecole d'Et{\'e} de Probabilit{\'e}s de Saint-Flour XXXIII - 2003}}}\ (\bibinfo  {publisher} {Springer},\ \bibinfo {year} {2007})\BibitemShut {NoStop}%
\bibitem [{\citenamefont {Liu}\ and\ \citenamefont {Liu}(2026)}]{liu2026}%
  \BibitemOpen
  \bibfield  {author} {\bibinfo {author} {\bibfnamefont {Z.}~\bibnamefont {Liu}}\ and\ \bibinfo {author} {\bibfnamefont {Z.-W.}\ \bibnamefont {Liu}},\ }\href {https://arxiv.org/abs/2609.03944} {\bibinfo {title} {On the geometry, typicality, and detectability of quantum magic}} (\bibinfo {year} {2026}),\ \Eprint {https://arxiv.org/abs/2609.03944} {arXiv:2609.03944 [quant-ph]} \BibitemShut {NoStop}%
\bibitem [{\citenamefont {Akibue}\ \emph {et~al.}(2026)\citenamefont {Akibue}, \citenamefont {Raymond}, \citenamefont {Tamaki},\ and\ \citenamefont {Teramoto}}]{akibue2026}%
  \BibitemOpen
  \bibfield  {author} {\bibinfo {author} {\bibfnamefont {S.}~\bibnamefont {Akibue}}, \bibinfo {author} {\bibfnamefont {R.}~\bibnamefont {Raymond}}, \bibinfo {author} {\bibfnamefont {S.}~\bibnamefont {Tamaki}},\ and\ \bibinfo {author} {\bibfnamefont {K.}~\bibnamefont {Teramoto}},\ }\href {https://arxiv.org/abs/2606.12700} {\bibinfo {title} {Measurement geometry for quantum random access codes: Beyond nayak bound and toward optimality}} (\bibinfo {year} {2026}),\ \Eprint {https://arxiv.org/abs/2606.12700} {arXiv:2606.12700 [quant-ph]} \BibitemShut {NoStop}%
\bibitem [{\citenamefont {{trureturing}}(nd)}]{trureturingPauliRefutation}%
  \BibitemOpen
  \bibfield  {author} {\bibinfo {author} {\bibnamefont {{trureturing}}},\ }\href@noop {} {\bibinfo {title} {The signed pauli sum bound fails for three qubits}},\ \bibinfo {howpublished} {\url{https://the-omega-institute.github.io/trureturing-mdbook/Blueprint/D5/S3/Quantum/Information/SignedPauliSumNormRefutation.html}} (\bibinfo {year} {n.d.}),\ \bibinfo {note} {machine-checked Lean formalization. Accessed November 18, 2026}\BibitemShut {NoStop}%
\bibitem [{\citenamefont {Turkeshi}\ \emph {et~al.}(2025)\citenamefont {Turkeshi}, \citenamefont {Dymarsky},\ and\ \citenamefont {Sierant}}]{Turkeshi_2025}%
  \BibitemOpen
  \bibfield  {author} {\bibinfo {author} {\bibfnamefont {X.}~\bibnamefont {Turkeshi}}, \bibinfo {author} {\bibfnamefont {A.}~\bibnamefont {Dymarsky}},\ and\ \bibinfo {author} {\bibfnamefont {P.}~\bibnamefont {Sierant}},\ }\bibfield  {title} {\bibinfo {title} {Pauli spectrum and nonstabilizerness of typical quantum many-body states},\ }\href {https://doi.org/10.1103/PhysRevB.111.054301} {\bibfield  {journal} {\bibinfo  {journal} {Phys. Rev. B}\ }\textbf {\bibinfo {volume} {111}},\ \bibinfo {pages} {054301} (\bibinfo {year} {2025})}\BibitemShut {NoStop}%
\bibitem [{\citenamefont {Erd{\'e}lyi}\ and\ \citenamefont {Tricomi}(1951)}]{Erdelyi_1951}%
  \BibitemOpen
  \bibfield  {author} {\bibinfo {author} {\bibfnamefont {A.}~\bibnamefont {Erd{\'e}lyi}}\ and\ \bibinfo {author} {\bibfnamefont {F.~G.}\ \bibnamefont {Tricomi}},\ }\bibfield  {title} {\bibinfo {title} {{The asymptotic expansion of a ratio of gamma functions.}},\ }\href@noop {} {\bibfield  {journal} {\bibinfo  {journal} {Pacific Journal of Mathematics}\ }\textbf {\bibinfo {volume} {1}},\ \bibinfo {pages} {133 } (\bibinfo {year} {1951})}\BibitemShut {NoStop}%
\end{thebibliography}%

\end{document}